\documentclass[a4paper,fleqn]{cas-sc}

\usepackage[authoryear]{natbib}
\usepackage[utf8]{inputenc}
\usepackage{amsmath}
\usepackage{amssymb}
\usepackage{mathtools}
\usepackage{hyperref}
\usepackage{graphicx}
\usepackage{tabularx}
\usepackage{enumerate}
\usepackage{ragged2e}
\usepackage{float}
\usepackage{color,soul}
\usepackage{chngcntr}
\usepackage{caption}
\usepackage{subcaption}
\counterwithin{figure}{section}
\counterwithin{table}{section}
\usepackage{lscape}
\usepackage{wasysym}
\usepackage{upgreek}

\def\tsc#1{\csdef{#1}{\textsc{\lowercase{#1}}\xspace}}
\tsc{WGM}
\tsc{QE}
\tsc{EP}
\tsc{PMS}
\tsc{BEC}
\tsc{DE}

\begin{document}
\let\WriteBookmarks\relax
\def\floatpagepagefraction{1}
\def\textpagefraction{.001}

\shorttitle{Self-similarity and growth of non-linear MRTI --- Role of the magnetic field}

\shortauthors{M. T. Kalluri and A. Hillier}

\title [mode = title]{Self-similarity and growth of non-linear magnetic Rayleigh-Taylor instability --- Role of the magnetic field strength}                  

\author{Manohar Teja Kalluri}[
orcid=https://orcid.org/0000-0002-5441-9224
]
\ead{manohartejakalluri25@gmail.com}

\credit{Data curation, Formal analysis, Investigation, Methodology, Software, validation, Visualization, Writing —original draft, review and editing}
\affiliation{organization={Department of Mathematics and Statistics, University of Exeter},
    addressline={Laver Building, 23 North Park Road}, 
    city={Exeter},
    postcode={EX4 4QE}, 
    country={The United Kingdom}}

\author{Andrew Hillier}
\ead{a.s.hillier@exeter.ac.uk}
\credit{Conceptualization, Funding acquisition, Project administration, Resources, Supervision, Writing—reviewing}

\begin{abstract}
The non-linear regime of the magnetic Rayleigh-Taylor instability (MRTI) has been studied in the context of several laboratory and astrophysical systems. Yet, several fundamental aspects remain unclear. One of them is the self-similar evolution of the instability. Studies have assumed that non-linear MRTI has a self-similar, quadratic growth similar to hydrodynamic (HD) RTI. However, neither self-similarity nor quadratic growth has been proved analytically. Furthermore, an explicit understanding of the factors that control the growth of non-linear instability remains unclear. Magnetic fields are known to play a crucial role in the evolution of the instability. Yet, a systematic study discussing how the magnetic field influences the instability growth is missing. These issues were addressed by performing an analytical and numerical study of the MRTI with a uniform magnetic field. Our study reveals that the imposed magnetic field does not conform to the HD self-similar evolution. However, the influence of the imposed magnetic field decays with time ($t$) as $1/t$ relative to the other non-linear terms, making the MRTI conform to the HD self-similarity. Thus, the HD RTI self-similar scaling becomes relevant to MRTI at late times, when nonlinear dynamics dominate. Based on energy conservation, an equation for the mixing layer height ($h$) is derived, which demonstrates the quadratic growth of $h$ in time. This gave insight into various factors that could influence the non-linear growth of the instability. By studying MRTI at different  magnetic field strengths, we demonstrate the role of magnetic field strength on the nonlinear growth of MRTI. Thus, the current study analytically and numerically proves the role of magnetic fields on the evolution of MRTI.
\end{abstract}

\begin{highlights}
\item Self-similarity of magnetic Rayleigh-Taylor instability (MRTI) was evidenced analytically and numerically.
\item Quadratic growth of mixing layer height for non-linear MRTI was derived analytically.
\item A formula for non-linear growth constant ($\alpha_{mhd}$) was derived. The formula highlights that energy dissipation, energy partition, energy anisotropy could dictate the non-linear growth constant.
\item The role of the magnetic field and the influence of magnetic field strength on the $\alpha_{mhd}$, energy dissipation, energy partition, energy anisotropy were elucidated.
\end{highlights}

% Keywords
\begin{keywords}
Rayleigh-Taylor instability \sep Magnetic fields \sep self-similarity \sep non-linear instability
\end{keywords}

\maketitle

\section{Introduction} \label{intro}

\begin{figure}
    \centering
    \includegraphics[width = \textwidth]{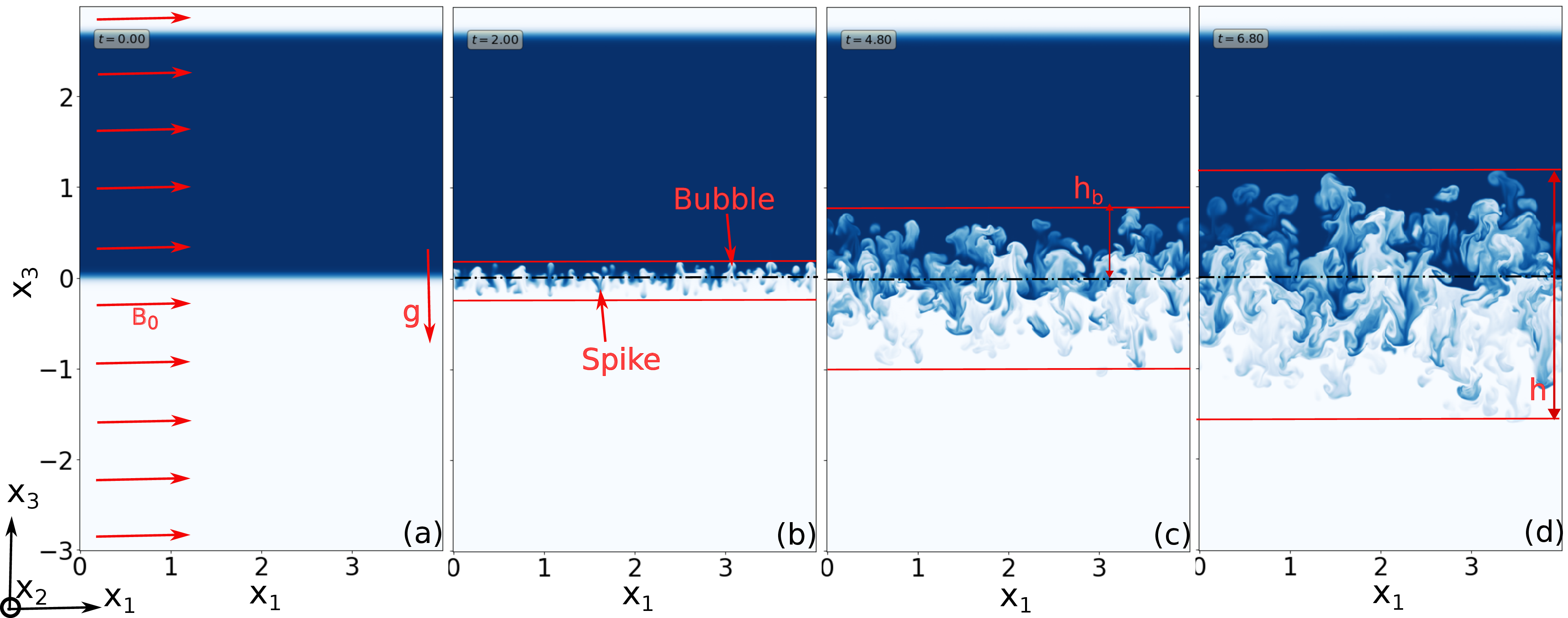}
    \caption{Figure showing the initial configuration (a), and evolution (b, c, d) of magnetic Rayleigh-Taylor instability mixing layer through the density contours (2D slice at mid $x_2$-plane) at different time instants $t {=} 2.0,$ $t {=} 4.80,$ $t {=} 6.80$ \textit{(from left to right)}. The snapshots correspond to $B_0 = 5\% B_c$ case. The red lines mark the boundaries of the mixing layer. The dash-dotted black line is the center line $x_3 {=} 0$. The distance between the red lines is the height of the mixing layer.}
    \label{MRTI_config}
\end{figure}

When a fluid of high-density ($\rho_h$) is supported by a fluid of low-density ($\rho_l$) in the presence of gravity, an infinitesimal perturbation at the interface of the two fluids could lead to the penetration of one fluid into the other (see figure \ref{MRTI_config}(a)). This phenomenon is called the \textit{Rayleigh-Taylor instability (RTI)} \citep{Rayleigh_1883, Taylor1950, Zhou_2024book}. The structures of low-density fluid penetrating the high-density fluid are called bubbles, and vice versa are called spikes (see figure \ref{MRTI_config}(b)). 

The hydrodynamic (HD) RTI evolves in three phases \citep{birkhoff1954taylor}. In the initial \textit{linear phase}, mixing layer grows exponentially with time $(t)$ i.e., $h \propto e^{\sqrt{\mathcal{A}kg} t}$ \citep{Taylor1950}. Here, $g$ is the acceleration due to gravity, $k$ is the perturbation wave mode, and the Atwood number $(\mathcal{A})$ represents the non-dimensionalized density difference between the two fluids, defined as $ \mathcal{A} = \frac{\rho_h {-} \rho_l}{\rho_h {+} \rho_l}$. Following the linear phase, the instability enters the nonlinear regime through a brief transitional regime, where the system is not necessarily turbulent. During the transitional phase, the system is nonlinear but may not be turbulent, with the shearing of plumes as they penetrate through the other fluid. Beyond this, as the instability evolves, the mixing layer becomes turbulent. The system only attains a self-similar state eventually as the instability evolves. In this self-similar phase, the height ($h$) of the mixing layer (region of penetration or mixing of the two fluids) \textit{grows quadratically} in time \citep{birkhoff1954taylor, birkhoff1955taylor}. The equation for the quadratic growth of the mixing layer height
\begin{equation}
    h = \alpha_{hd} \mathcal{A} g t^2 + 2 \sqrt{\alpha_{hd} \mathcal{A} g h_0} t + h_0
    \label{HD_h}
\end{equation}
was derived analytically in the \textit{Boussinesq limit} by \cite{ristorcelli_clark_2004}. Here, $\alpha_{hd}$ is the non-linear growth constant, and $h_0$ is the height of mixing layer at time $t {=} 0$, assuming equation \ref{HD_h} is obeyed from $t{=}0$. The above equation was also verified numerically \citep{ristorcelli_clark_2004} and experimentally \citep{dalziel_1999, Linden1991} in the Boussinesq limit.

Following the dimensional analysis by \citet{Fermi1953}, the quadratic growth is expected to be valid for the non-Boussinesq case. The quadratic growth in this limit was studied numerically by \citet{Youngs1991, Cabot2006, Cabot_2013, Dimonte2004}. Thus, the mixing layer of non-linear HD RTI is reported to grow in a self-similar fashion $(h \approx \alpha_{hd} \mathcal{A} gt^2, t \gg 1)$ for all Atwood numbers. $\alpha_{hd}$ represents how fast the mixing layer grows. Typically, $\alpha_{hd}$ is calculated from the slope of $h$ and $t^2$ or $(\partial h/\partial t)^2$ and $h$. The experimental and numerical studies of the HD RTI showed distinct $\alpha_{hd}$ values, with a wide error range. These disagreements are summarized in \citet{GLIMM2001}. Following previous studies, $\alpha_{hd}$ (based on $h_b$) is expected to be between 0.03 and 0.08.

\begin{figure}
     \centering
     \begin{subfigure}[b]{0.49\textwidth}
         \centering
         \includegraphics[width=\textwidth]{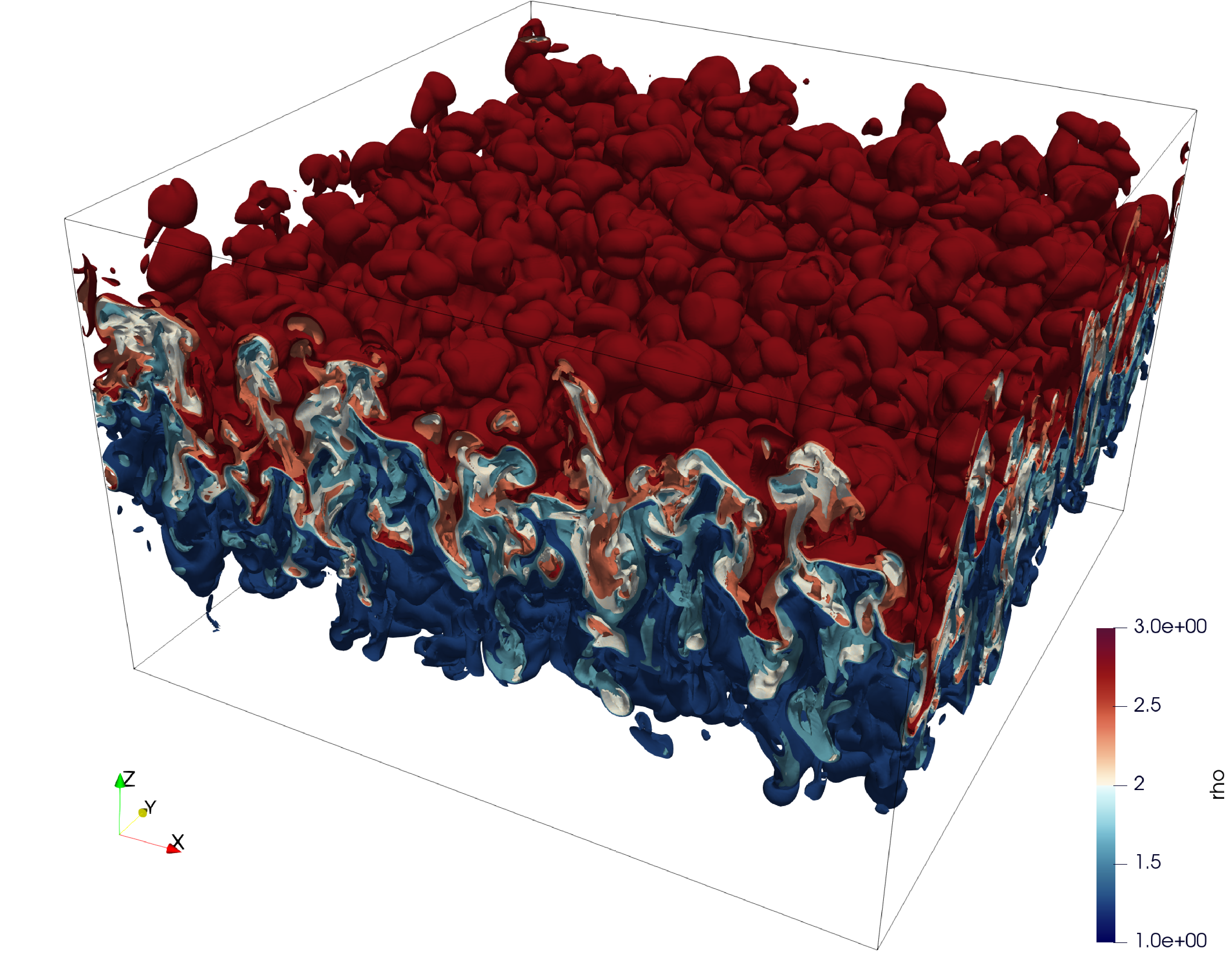}
     \end{subfigure}
     \hfill
     \begin{subfigure}[b]{0.49\textwidth}
         \centering
         \includegraphics[width=\textwidth]{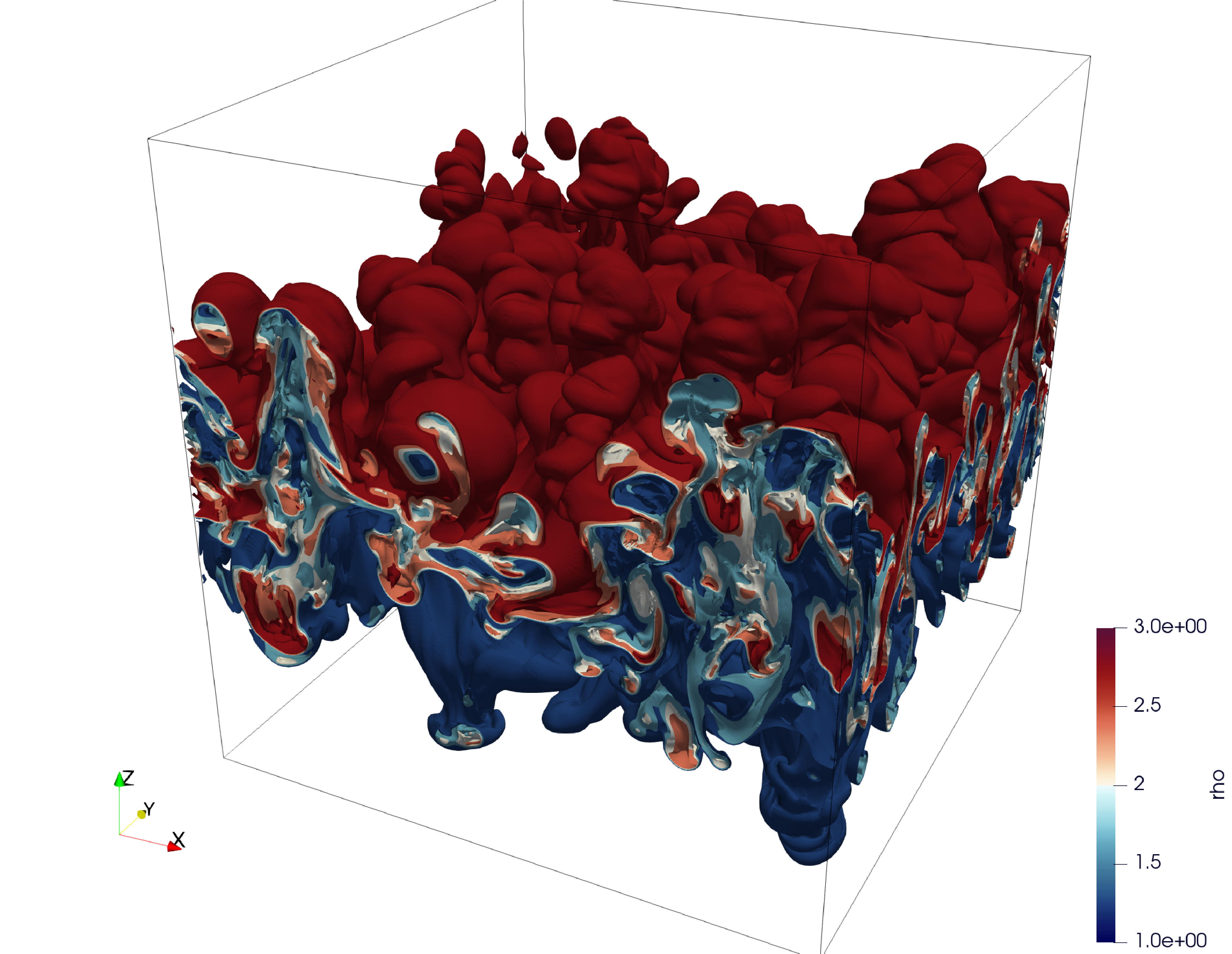}
     \end{subfigure}
     \caption{Instantaneous density contours of MRTI at two different magnetic field strengths: \textit{(left)} $B_0 = 5\% B_c$; \textit{(right)} $B_0 = 25\% B_c$. Both the contours are plotted at the same time instant $t = 7$.}
     \label{isocontours}
\end{figure}

A special case of the RTI evolution in the presence of magnetic fields is the \textit{magnetic Rayleigh-Taylor instability} (MRTI). MRTI is typically seen in fusion energy systems \citep{Zhou2024}, geophysical and astrophysical systems \citep{ZHOU2021, ZHOU2017_1, ZHOU2017_2}. Assuming the approximate flux frozen condition (strong coupling of magnetic field lines and streamlines), the instability evolution demands deformation of magnetic field lines. This deformation develops magnetic tension that resists the growth of instability. The suppression of instability by the magnetic field is reflected in the linear growth rate ($\sigma$) equation of an incompressible MRTI \citep{Chandrasekhar_1961},
\begin{equation}
    \sigma_{mhd} = \sqrt{\mathcal{A} k g {-} \frac{2 k^2 B^2 cos^2 \theta}{(\rho_h {+} \rho_l)}},
    \label{Growthrate_MRT}
\end{equation}
where $\theta$ is the angle between $\textbf{k}$ and $\textbf{B}$. Note that the wave vector and imposed magnetic field must be in the same horizontal plane, tangent to the interface. From equation \ref{Growthrate_MRT}, the suppression effect is prominent on the wave modes parallel to the field ($\theta {=} 0$, called undular modes), and the effect diminishes with increasing misalignment ($\theta$). Perturbations perpendicular to the magnetic field ($\theta {=} \pi/2$, called interchange modes) experience no suppression and grow similar to HD RTI. From equation \ref{Growthrate_MRT}, for a given $\mathcal{A}, g, \theta$, $B$, the growth rate decreases with increasing $k$. The value of $k$ at which $\sigma_{mhd} = 0$ is called the critical wave number $(k_c)$, given by
\begin{equation}
    k_c {=} \frac{(\rho_h - \rho_l) g}{2 B^2 cos^2 \theta}.
    \label{kc}
\end{equation}
While wave modes smaller than $k_c$ continue to grow, modes larger than $k_c$ are suppressed by the magnetic field. Thus, the magnetic field \textit{selectively suppresses} the perturbations. Thus, unlike the HD RTI, MRTI does not develop for every perturbation, but grows only when the perturbation includes wave modes smaller than the critical wave mode. The suppression of wave modes by the magnetic field has an important consequence on the nature of non-linear MRTI. In the weak magnetic field case, where a wide range of wave modes grow, the mixing layer is turbulent. As we increase the magnetic field strength, $k_c$ decreases. This results in the mixing layer with increasingly large-scale plumes at strong magnetic field strengths \citep{Jun1995, Stone2007a, Carlyle2017}. This is also found in our simulations, see Figure \ref{isocontours} where we show density iso-contours of two different magnetic field strengths at the same time instant.

Similarly, for a given $\mathcal{A}, g, \theta$ and $k$, the linear growth rate decreases with increasing $B$. The value of $B$ at which $\sigma_{mhd} = 0$ is called the critical magnetic field strength ($B_c$), given by
\begin{equation}
    B_c {=} \sqrt{\frac{(\rho_h - \rho_l) g}{2 k cos^2 \theta}}.
    \label{Bc}
\end{equation}
$B_c$ is originally defined for a single-mode undular ($\theta = 0$) perturbation. For a multi-mode system, $B_c$ is different for each $k$. Note that $B_c \rightarrow \infty$ when the magnetic field is perpendicular to all wave modes, i.e., no finite amount of magnetic field strength can suppress these wave modes.

In the context of non-linear MRTI mixing layer growth, studies so far \citep{Jun1995, Stone2007a, Carlyle2017} assumed that $h$ grows quadratically in time ($h \approx \alpha_{mhd} \mathcal{A} g t^2$, $ \alpha_{mhd}$ is the non-linear growth constant), similar to the HD RTI (equation \ref{HD_h}). \citet{Stone2007a} performed a parametric study with field strength increasing from $0$ to $60 \% B_c$ and reported that $\alpha_{mhd} (= h/\mathcal{A} g t^2)$ increases from $0.021$ (HD) to $0.034$ ($60 \% B_c$).

There are three caveats in the existing non-linear MRTI theory. First, the \textit{self-similar growth} of the MRTI mixing layer is an \textit{assumption}, but not proven analytically. Such an assumption is also not obvious, because the linear growth rate equation (equation \ref{Growthrate_MRT}) shows that the gravity and magnetic field act at different length scales, $k$ and $k^2$ respectively. This hints that the magnetic field might follow a different scaling compared to gravity, consequently deviating the MRTI system from the HD self-similarity. Thus, the assumption of self-similarity is questionable.

Second, the assumption of $h \approx \alpha_{mhd} \mathcal{A} g t^2$ is not backed by an \textit{analytical derivation} of $h$ for the MRTI case. Further, the assumption considers only the gravitation force and does not reflect the influence of the magnetic force on the mixing layer growth. In the non-linear HD RTI growth, there is only one force acting on the system, gravity. By dimensional analysis, it is reasonable to suppose that $h \propto \mathcal{A}gt^2$. But in the MRTI, there are two competing forces --- gravity and magnetic tension. While the gravity drives the instability, the magnetic field suppresses it, see equation \ref{Growthrate_MRT}. Hence, the assumption $h \propto \mathcal{A}gt^2$ with only the gravity term is dubious for the MRTI, unless evidenced. 

Third, supposing the argument $h \approx \alpha_{mhd} \mathcal{A}gt^2$ is correct and magnetic field influences the growth through $\alpha_{mhd}$ (as reported by \cite{Stone2007a, Carlyle2017}), it is not clear how the magnetic field influences $\alpha_{mhd}$. This is due to the lack of an explicit understanding of what factors control the non-linear growth constant in both HD and magnetic field cases.

Targeting the issues, we first perform an analytical analysis to validate the self-similarity assumption and the appropriateness of HD RTI scaling for MRTI in $\S$\ref{self-similarity}. Later, an analytical equation for the temporal growth of the mixing layer will be derived in $\S$\ref{h-derivation}. The equation of mixing layer height sheds light on the factors that control the instability growth. In $\S$\ref{Results}, we validate the conclusions of $\S$\ref{self-similarity} and $\S$\ref{h-derivation} through numerical simulations at different magnetic field strengths. The methodology for the numerical simulations is presented in $\S$\ref{Methodology}.

\section{Analytical investigation of hydrodynamic self-similarity in MRTI} \label{self-similarity}

Following the numerous non-linear HD RTI studies, the non-linear MRTI studies so far \citep{Jun1995, Stone2007a, Stone_2007b, Carlyle2017} \textit{assumed} that the mixing layer has a self-similar quadratic growth. However, \textit{an analytical proof of the self-similar evolution and the relevance of HD scaling for the non-linear MRTI is lacking}. To verify this assumption, we perform an analytical self-similar analysis of the ideal MHD equations with the HD scaling. We stress that this study aims to validate the relevance of HD RTI scaling to MRTI.

Before we look at the present self-similar analysis, we will first discuss the approach of the previous HD RTI studies. The governing equations for the variable density HD RTI are \citep{COOK_DIMOTAKIS_2001}:
\begin{subequations}
    \begin{align}
        \partial_t \rho \mathbf{U} {+} (\mathbf{U} \cdot \mathbf{\nabla}) \rho \mathbf{U}  & = {-} \mathbf{\nabla} P_h - \rho \mathbf{g} + \frac{1}{Re} \left( \mathbf{\nabla^2 \mathbf{U}} + \frac{1}{3} \mathbf{\nabla}(\mathbf{\nabla} \cdot \mathbf{U}) \right), \text{ (Momentum equation)} \label{CookNS} \\
        \partial_t \rho {+} (\mathbf{U} \cdot \mathbf{\nabla}) \rho & =  \frac{1}{Re Sc} \left( \nabla^2 \rho - \frac{1}{\rho} (\nabla \rho)^2 \right), \text{ (Mass conservation equation)} \label{Cookdensity} \\
        \mathbf{\nabla} {\cdot}  \mathbf{U} & = -\frac{1}{Re Sc} \nabla \cdot \left(\frac{1}{\rho} \nabla\rho \right). \text{ (Velocity divergence condition)} \label{Cookdivu}
    \end{align}
    \label{Cook2001}
\end{subequations}
$\textbf{U}$ is the instantaneous flow velocity, $\rho$ is the fluid density, $P_h$ is the total gas pressure. Here on, the vector variables are written in bold. Gravity is assumed to act along the $x_3$ direction (i.e., $\textbf{g} {=} (0, 0, -g)$). Thus, the MRTI grows along $x_3$, the direction of statistical inhomogeneity. The other two directions $x_1,$ $x_2$ are statistically homogeneous. $Re$ is the Reynolds number, defined as $Re = U L/\nu$ where $U, L$ are characteristic velocity and length scale, and $\nu$ is the fluid viscosity. The Schmidt number ($Sc$) is defined as the ratio of fluid viscosity to mass diffusivity ($\mathcal{D}$) (i.e., $Sc = \nu/\mathcal{D}$). 

Note that, despite being an incompressible system, the velocity divergence is non-zero. This is due to the Fickian modelling of the mass diffusion, where the diffusion term of mass continuity equation is considered as $\nabla \cdot \left( \rho \frac{1}{Re Sc} \nabla \mathcal{C}_i \right)$, $\mathcal{C}_i$ being the local mass fraction of the fluid $i$. Using the mass continuity equation for the two fluids, we can eliminate the $C_i$ using the condition $C_1 + C_2 = 1$. Thus, we end up with the mass continuity equation of form $\frac{\mathrm{D} \rho}{\mathrm{D} t} = \rho \nabla \cdot \left( \frac{1}{Re Sc}  \frac{1}{\rho} \nabla \rho \right)$. Comparing the above equation with $\frac{\mathrm{D} \rho}{\mathrm{D} t} = - \rho (\nabla \cdot u)$, we get equation \ref{Cookdivu}. A detailed derivation can be seen in \citet{Sandoval1995}.

The inclusion of a magnetic field introduces a new body force, ($(\nabla \times \textbf{B}) \times \textbf{B}$), in the momentum equation. Note that $\textbf{B}$ is normalized by the square root of the magnetic permeability of free space ($\sqrt{\mu_0}$). Rewriting the new body force term as the sum of magnetic pressure, $\nabla \left(\frac{\textbf{B} \cdot \textbf{B}}{2} \right)$, and magnetic tension, $(\textbf{B} \cdot \nabla) \textbf{B}$. The MHD momentum equation becomes
\begin{equation}
    \rho \partial_t \mathbf{U} {+} \rho (\mathbf{U} \cdot \mathbf{\nabla}) \mathbf{U}  = {-} \mathbf{\nabla} P {+} (\textbf{B} \cdot \nabla) \textbf{B} - \rho \mathbf{g} + \frac{1}{Re} \left( \mathbf{\nabla^2 \mathbf{U}} + \frac{1}{3} \mathbf{\nabla}(\mathbf{\nabla} \cdot \mathbf{U}) \right)
\end{equation}
$P$ is the sum of gas and magnetic pressures $\left(P_h + \frac{\textbf{B} \cdot \textbf{B}}{2}\right)$. In the initial quiescent state ($\textbf{U} = 0$), the above equation reduces to $\nabla P_0 = -\rho_0 g$, the hydrostatic condition. The pressure and gravity terms can be split into initial and fluctuating quantities based on Reynolds decomposition as $\nabla P = \nabla P_0 + \nabla p$ and $\rho\mathbf{g} = \rho_0 \mathbf{g} + \delta \rho \mathbf{g}$. Rewriting the pressure and gravity terms as above and substituting the hydrostatic condition, we reduce the above momentum equation to
\begin{equation}
    \rho \partial_t \textbf{U} {+} \rho (\textbf{U} \cdot \nabla) \textbf{U} = - \nabla p {+} (\textbf{B} \cdot \nabla) \textbf{B} - \delta \rho \textbf{g} + \frac{1}{Re} \left( \mathbf{\nabla^2 \mathbf{U}} + \frac{1}{3} \mathbf{\nabla}(\mathbf{\nabla} \cdot \mathbf{U}) \right),
\end{equation}
where $\delta \rho$ and $p$ are fluctuating quantities about initial quantities.

In the context of large-scale self-similar evolution of instability, the diffusion terms play an insignificant role. Therefore, we consider the ideal MHD governing equations (equations \ref{Cook2001} and the induction equation, which governs the evolution of the magnetic field, with Laplacian terms neglected in all of them). The ideal MHD governing equations become
\begin{subequations}
    \begin{align}
        \rho \partial_t \textbf{U} {+} \rho (\textbf{U} \cdot \nabla) \textbf{U} & = - \nabla p {+} (\textbf{B} \cdot \nabla) \textbf{B} - \delta \rho \textbf{g}, \\
        \partial_t \textbf{B} {+} (\textbf{U} \cdot\nabla) \textbf{B} & = (\textbf{B} \cdot \nabla) \textbf{U}, \text{  (Induction equation)} \label{Inductioneqn} \\
        \partial_t \rho {+} (\textbf{U} \cdot \nabla) \rho & = 0, \\
        \nabla \cdot \textbf{U} = \nabla \cdot \textbf{B} & = 0.
    \end{align}
    \label{governing_eqns}
\end{subequations}

The self-similar analysis for a HD Boussinesq case ($\nabla \rho \rightarrow 0$) was performed by \citet{ristorcelli_clark_2004}. In this study, \citet{ristorcelli_clark_2004} obtained the first and second moment equations averaged along the homogeneous directions. Assuming self-similar solutions of the form $\Phi(x_3, t) {=} \phi_t(t) \phi_{\eta}(\eta),$ $ \eta {=} x_3/h$, the first and second moment equations are written in terms of $\Phi$. Following this, four terms are considered: \\ 
$(i)$ the advection term of the first-order density equation, $\langle u_3 \partial \rho/ \partial x_3 \rangle$; \\
$(ii)$ the production term of the $\langle \rho \rho \rangle$ equation, $\langle u_3 \rho \rangle \partial \rho_0/ \partial x_3 $; and \\
$(iii), (iv)$ the two production terms of the $\langle u_3 \rho \rangle$ equation, $\langle (\delta \rho)^2 g_i \updelta_{i3} \rangle$ and $\langle u_3 u_3 \partial \rho/ \partial x_3 \rangle$. \\
Using these four terms, \citet{ristorcelli_clark_2004} obtain an ordinary differential equation (ODE) for $\mathrm{d}h/\mathrm{d}t$ in terms of $h$. The ODE is solved to obtain a quadratic equation for $h$. 

Implementing the above methodology for the MRTI case, we derived the first and second order moments of equations \ref{governing_eqns} in terms of the self-similar solutions $\Phi(\eta, t)$. However, this did not lead to insights different from the HD case. This is because the terms listed above are independent of the magnetic field effects, and we end up with the same ODE as obtained for the HD RTI case. That is, the influence of the magnetic field on the mixing layer height remains obscure. Comparing other terms also did not demonstrate the simultaneous effect of gravity and magnetic fields since a term coupling $h$, $g$, and $\textbf{B}$ is absent. Thus, the method of \cite{ristorcelli_clark_2004} was found unsuitable for the MRTI case, with two competing forces, gravity and magnetic tension. This necessitated a different approach. 

Alternative to \citet{ristorcelli_clark_2004}'s method, we use the \textit{proof by contradiction} method to determine the relevance of the HD RTI scaling to the MRTI. The idea is to assume that the MRTI mixing layer is self-similar and obeys HD scaling laws, i.e., $h$ grows quadratically. The flow variables will be non-dimensionalized with the HD RTI scaling. A self-similar variable ($\xi$) relating the non-homogeneous spatial parameter ($x_3$) and the time ($t$) will be formed. If the non-dimensionalized MHD governing equation reduces to ODEs, it signifies that the MRTI can be self-similar and has quadratic growth. Else, we conclude that the MRTI does not accept the HD self-similar solutions. Further, this method allows us to see which terms disagree with the HD self-similarity scaling. Thus, this method can validate the self-similarity and the quadratic growth of MRTI.

Towards the plan proposed above, consider the \textit{an approximate variable-density ideal MHD equations}, equations \ref{governing_eqns}. We consider the non-Boussinessq case ($\nabla \rho > 0$). Splitting the flow variables $(\textbf{U}, \textbf{B})$ into the initial ($\textbf{U}_0, \textbf{B}_0$) and fluctuating components ($\textbf{u}, \textbf{b}$). Consider the case of an initial stationary system with uniform magnetic field in $x_1$, the equations \ref{governing_eqns} become:
\begin{subequations}
    \begin{align}
        \partial_t \rho u_i {+} \partial_j (\rho u_j u_i) & = - \partial_i p {+} \partial_j (B_{0j} b_i) {+} \partial_j (b_j b_i) {-} \delta \rho g \updelta_{i3}, 
        \label{momentum_eqn} \\
        \partial_t b_i {+} \partial_j (u_j b_i) & = \partial_j (B_{0j} u_i) {+} \partial_j (b_j u_i), 
        \label{induction_eqn} \\
        \partial_t \rho {+} \partial_j (u_j \rho) & = 0,
        \label{mass_equation} \\
        \partial_i u_i = \partial_i b_i & = 0.
        \label{incompressibility} 
    \end{align}
    \label{governing_eqns_2}
\end{subequations}
The above equations are written in tensor notation, where $\{i, j\} \in [1,3]$, $\updelta_{i3}$ represents the Kronecker delta function. The Einstein summation rule is applicable everywhere. Here on, $\partial_j \star$ represent partial derivative of $\star$ with respect to $x_j$. The above equations are obtained using the condition \ref{incompressibility}. $\partial_j \mathbf{B_0} {=} 0$ and $\partial_t \mathbf{B_0} {=} 0$ as $\mathbf{B_0}$ is uniform and constant. 

Since the aim is to validate the relevance of the HD scaling to the MHD case, we \textit{non-dimensionalise the flow quantities in terms of the HD scaling}. The height of the HD mixing layer at late time can be approximated to $h {\propto} \mathcal{A} g t^2$ cf., equation \ref{HD_h}. The characteristic speed at the boundaries of the mixing layer is proportional to the temporal growth rate of the mixing layer height, $u \propto \partial_t h \propto \mathcal{A} g t$. Hence, we choose $\mathcal{A} gt$ as the characteristic speed $(u_c)$ for non-dimensionalisation. From the definition of Alfven velocity $(v_A = B/\sqrt{\rho})$, the magnetic field can be written in terms of velocity and density, so we choose $u_c \sqrt{\rho_m}$ as the non-dimensional parameter for magnetic field. $\rho_m$ is the non-dimensional quantity for the density, and is the arithmetic mean of $\rho_h$ and $\rho_l$. The pressure is non-dimensionalised as $\rho_m u_c^2$. The flow parameters can be written in non-dimensionalised forms as shown below:
\begin{align*}
    \rho {=} \rho_m \Tilde{\rho}, \hspace{10pt} u {=} \mathcal{A} g t \Tilde{u}, \hspace{10pt} b {=} \sqrt{\rho_m} \mathcal{A} g t \Tilde{b}, \hspace{10pt} p {=} \rho_m (\mathcal{A} g t)^2 \Tilde{p}.
\end{align*}
$\Tilde{\star}$ represent the non-dimensional form of variable $\star$. Rewriting the equations \ref{momentum_eqn}, \ref{induction_eqn}, \ref{mass_equation} in the non-dimensional form and averaging along the homogeneous directions $\left( \langle \star \rangle {=} \frac{1}{L_{x_1} L_{x_2}} \int_{0}^{L_{x_1}} \int_{0}^{L_{x_2}} \star \mathrm{d}x_1 \mathrm{d}x_2 \right)$, we get
\begin{subequations}
    \begin{align}
        \begin{split}
        t \partial_t \langle \Tilde{\rho} \Tilde{u}_i \rangle {+} \langle \Tilde{\rho} \Tilde{u}_i \rangle {+} \mathcal{A} g t^2 \partial_3 \langle (\Tilde{\rho} \Tilde{u}_3 \Tilde{u}_i) \rangle {+} \mathcal{A} g t^2 \partial_3 \langle \Tilde{p} \rangle {-} \mathcal{A} g t^2 \partial_3 \langle \Tilde{b}_3 \Tilde{b}_i \rangle  {+} \frac{1}{\mathcal{A}} \langle \delta \Tilde{\rho} \updelta_{i3} \rangle  & = 0, 
        \end{split} \\
        \partial_3 \langle \Tilde{u}_3 \Tilde{b}_i \rangle  {-} \partial_3 \langle \Tilde{b}_3 \Tilde{u}_i \rangle  & = 0, \\
        \partial_t \langle \Tilde{\rho} \rangle  {+} \mathcal{A} g t \partial_3 \langle \Tilde{u}_3 \Tilde{\rho} \rangle  & = 0.
    \end{align}
\end{subequations}
Note that, $\partial_{1} \langle \star \rangle {=} 0,$ $\partial_{2} \langle \star \rangle {=} 0,$ since $x_1$ and $x_2$ are directions of statistical homogeneity. 

The variables $x_3$ and $t$ can be scaled to get a self similar variable $\xi {=} \sqrt{\frac{x_3}{\mathcal{A}g}}\frac{1}{t}$. The partial derivatives of $x_3$ and $t$ can now be rewritten in terms of $\xi$ as $\partial_t = \frac{-\xi}{t} \partial_\xi,$ $\partial_{3} = \frac{\xi}{2 x_3} \partial_\xi $. The above system of equations is modified to
\begin{subequations}
\begin{align}
    \begin{split}
    {- \xi}\mathrm{d}_\xi \langle \Tilde{\rho} \Tilde{u}_i \rangle {+} \xi \langle \Tilde{\rho} \Tilde{u}_i \rangle {+} \frac{1}{2} \mathrm{d}_\xi \langle \Tilde{\rho} \Tilde{u}_3 \Tilde{u}_i \rangle {+} \frac{1}{2} \mathrm{d}_{\xi} \langle \Tilde{p} \rangle {-} \frac{1}{2} \mathrm{d}_\xi \langle \Tilde{b}_3 \Tilde{b}_i \rangle  {+} \xi \langle \frac{\delta \Tilde{\rho} \updelta_{i3}}{\mathcal{A}} \rangle  & = 0,
    \end{split} \\
    \mathrm{d}_\xi \langle \Tilde{u}_3 \Tilde{b}_i \rangle  {-} \mathrm{d}_\xi \langle \Tilde{b}_3 \Tilde{u}_i \rangle  & = 0, \\
    \xi^2 \mathrm{d}_\xi \langle \Tilde{\rho} \rangle  {+} \frac{1}{2} \mathrm{d}_\xi \langle \Tilde{u}_3 \Tilde{\rho} \rangle  & = 0.
    \label{low_order_eqns}
\end{align}
\end{subequations}
The solutions for the above ODEs are functions of $\xi$. % Thus, momentum, density, and fluctuating magnetic field can take self-similar solutions similar to the HD scaling. 

Next, we investigate the higher-order quantities, turbulent kinetic energy (TKE) and turbulent magnetic energy (TME).
To obtain the equation of TKE, defined as $\frac{1}{2} \langle \rho u_i u_i \rangle$, we multiply equation \ref{momentum_eqn} with $u_i$ and average along homogeneous directions to get

\begin{equation}
    \partial_t \left\langle \rho \frac{u_i u_i}{2} \right\rangle {+} \partial_3 \left\langle u_3 \rho \frac{u_i u_i}{2} \right\rangle  {+} \partial_3 \langle u_3 p \rangle  {-} B_{0} \langle u_i \partial_1 b_i \rangle {-} \langle u_i b_j \partial_j b_i \rangle  {+} \langle u_i \delta \rho g \updelta_{i3} \rangle  = 0.
    \label{TKE_eqn}
\end{equation}
In terms of non-dimensional quantities and self-similar variables, the equation \ref{TKE_eqn} becomes
\begin{equation}
    \begin{split}
     \xi \langle \Tilde{\rho} \Tilde{u}_i \Tilde{u}_i \rangle {-} \xi^2 \partial_\xi \left\langle  \Tilde{\rho} \frac{\Tilde{u}_i \Tilde{u}_i}{2} \right\rangle {+} \frac{1}{2} \partial_\xi \left\langle \Tilde{u}_3 \Tilde{\rho} \frac{\Tilde{u}_i \Tilde{u}_i}{2} \right\rangle {-} \frac{1}{2} \partial_\xi \langle \Tilde{u}_3 \Tilde{p} \rangle {-} \frac{1}{2 \sqrt{\rho_m} C_1 \mathcal{A} g t} B_0 \langle \Tilde{u}_i \partial_{\xi} \Tilde{b}_i \rangle {-} \frac{1}{2} \left \langle \Tilde{u}_i \partial_{\xi} \left ( \frac{\Tilde{b}_{1} \Tilde{b}_i}{C_1} {+} \frac{\Tilde{b}_{2} \Tilde{b}_i}{C_2} {+} \Tilde{b}_{3} \Tilde{b}_i \right) \right \rangle \\ {-} \xi \frac{1}{\mathcal{A}}  \langle \Tilde{\rho} \Tilde{u}_i \delta_{i3} \rangle  = 0,
    \end{split}
    \label{self_similar_TKE}
\end{equation}
The $\partial_1$, $\partial_2$ terms are written in terms of $\partial_3$ based on the scaling assumption $x_i {=} C_i x_3, i {=} [1, 2]$. This assumption is based on the hypothesis that the flow structures evolve with a fixed scaling in each direction. In other words, the system has a fixed anisotropy during the self-similar phase. This assumption will be verified later in $\S$\ref{section:verification}.

Similarly, the equation of TME, defined as $\frac{1}{2} \langle b_i b_i \rangle$, is obtained by multiplying equation \ref{induction_eqn} with $b_i$, and the resultant equation is averaged along the homogeneous directions to get
\begin{equation}
    \partial_t \left\langle \frac{b_i b_i}{2} \right\rangle {+} \partial_3 \left\langle u_3 \frac{b_i b_i}{2} \right\rangle {-} B_{0} \partial_1 \langle b_i u_i \rangle {+} B_{0} \langle u_i \partial_1 b_i \rangle {-} \langle b_j \partial_j (b_i u_i) \rangle {+} \langle u_i b_j \partial_j b_i \rangle = 0.
    \label{TME_eqn}
\end{equation}
Writing the equation \ref{TME_eqn} in terms of the non-dimensional variables and the self-similar variable, we get
\begin{equation}
    \begin{split}
         \xi \langle \Tilde{b}_i \Tilde{b}_i \rangle {-} \xi^2 \partial_\xi \left\langle \frac{\Tilde{b}_i \Tilde{b}_i}{2} \right\rangle {+} \frac{1}{2} \partial_\xi \left\langle \Tilde{u}_3 \frac{\Tilde{b}_i \Tilde{b}_i}{2} \right\rangle {+} \frac{1}{2} \left \langle \Tilde{u}_i \partial_{\xi} \left( \frac{\Tilde{b}_{1} \Tilde{b}_i}{C_1} {+} \frac{\Tilde{b}_{2} \Tilde{b}_i}{C_2} {+} \Tilde{b}_{3} \Tilde{u}_i \right) \right \rangle {+} \frac{1}{2 \sqrt{\rho_m} C_1 \mathcal{A} g t} B_{0} \langle \Tilde{u}_i \partial_{\xi} \Tilde{b}_i \rangle {+} \frac{1}{2} \partial_{\xi} \langle \Tilde{b}_3 \Tilde{b}_i \Tilde{u}_i \rangle = 0
    \end{split}
    \label{self_similar_TME}
\end{equation}

Unlike the mass, momentum, and induction equations, the TKE and TME equations do not reduce to ODEs in $\xi$ alone. This is due to the initial magnetic field term $\left( \frac{1}{2 \sqrt{\rho_m} C_1 \mathcal{A} g t} B_{0} \langle \Tilde{u}_i \partial_{\xi} \Tilde{b}_i \rangle \right)$ which varies with $t$ besides $\xi$. That is, the system is not self-similar unless the influence of the initial magnetic field term is diminished. However, in MRTI, we find that the temporal variation of the imposed magnetic field term varies as $1/t$ relative to the other non-linear terms like the gravitational potential energy (GPE). As the instability evolves, the influence of non-linear terms increases, and the influence of the imposed magnetic field decreases. Thus, at late time $t \gg 1$, the non-linear terms dominate the dynamics, leading the system towards self-similar evolution with scaling laws similar to the HD RTI. Note that no assumption on the density ratio was made in the derivation. Hence, the convergence towards self-similarity is expected to be true for both Boussinesq and non-Boussinesq cases. It is easy to see that the above equations (\ref{self_similar_TKE}, \ref{self_similar_TME}) reduce to self-similar form in the limit $B_0 = 0$.

To summarize, studies so far \textit{assumed} that the MRTI has a self-similar evolution, with the same scaling as HD RTI. To validate the assumption, we perform an analytical self-similar analysis, using the HD scaling for the ideal MHD equations. We find that the second-order equations do not reduce to ODEs. Hence, through proof by contradiction, we show that the MRTI evolution can be self-similar only when the non-linear terms dominate the imposed magnetic field. Nevertheless, with the inevitable domination of the non-linear dynamics term over the imposed magnetic field, it is expected that the MRTI can possess the same self-similarity scaling as the HD RTI. It is important to note that our approach demonstrates that the self-similar solutions of quadratic form are \textit{a} solution for the problem, but does not prove that they are \textit{the} only solution. % The $t$ scaling of velocity makes the analysis appropriate to buoyancy-driven flows like RTI.

\section{Expression for mixing layer height} \label{h-derivation}

Addressing the first issue discussed in $\S$\ref{intro}, $\S$\ref{self-similarity}, we expect that the MRTI can converge to HD self-similarity when the non-linear terms dominate the imposed magnetic field. Assuming the \textit{self-similar state is achieved}, in this section we address the other issues described in $\S$\ref{intro} --- analytic derivation of mixing layer height, lack of an understanding of the factors that control the non-linear MRTI growth. One of the ways to understand the non-linear MRTI growth is to derive an equation for the mixing layer height (similar to the equation \ref{HD_h}, since we know MRTI has scaling similar to HD RTI in the self-similar regime) and comparing the term that correspond to the non-linear growth constant ($\alpha_{mhd}$).

For the reasons described in $\S$\ref{self-similarity}, the method used by \cite{ristorcelli_clark_2004} to derive the mixing layer height equation is not useful in the context of MRTI. Further, the \cite{ristorcelli_clark_2004} method did not lead to an expression for $\alpha_{hd}$ in their study, making the approach unsuitable for the current goals, determining the physical process that controls the nonlinear growth constant. Hence, we propose a different, more fundamental approach to derive the mixing layer height \textit{in the self-similar regime} of MRTI. From the conservation of energy, we know that the GPE released is the sum of turbulent kinetic energy (TKE), turbulent magnetic energy (TME), and the energy dissipated up until the time, i.e.,
\begin{equation}
    -\underbrace{\int_V \delta \rho g x_3 \mathrm{d}V }_\text{GPE released} = \underbrace{\int_V \frac{1}{2} \rho u^2 \mathrm{d}V}_\text{TKE} {+} \underbrace{\int_V \frac{1}{2} b^2 \mathrm{d}V}_\text{TME} {+} \underbrace{D_E}_\text{Total energy dissipated}.
\end{equation}
This is similar to the energy conservation technique proposed by \citet{Linden_Redondo_Youngs_1994} for HD RTI, except for an extra quantity, TME, due to the inclusion of magnetic fields. The equation of energy dissipation will depend on the choice governing equations, velocity divergence condition, and other assumptions. The total energy dissipation for the current study is mentioned in $\S$\ref{section:verification}. The current analysis is independent of how energy dissipation is calculated. Hence, an expression for $D_E$ is not presented here.

The HD RTI study of \cite{Youngs1991} showed that the energy dissipation is proportional to the TKE in the system. Extending it to the MRTI case, we propose that the total energy dissipation $D_E$ in the MHD system is equal to the sum of energy dissipation due to TKE and TME. Let $D_E = C_{diss} \left(\int_V \frac{1}{2} \rho u^2 \mathrm{d}V {+} \int_V \frac{1}{2} b^2 \mathrm{d}V \right)$. Hence, the above equation reduces to
\begin{equation}
    - \int_V \delta \rho g x_3 \mathrm{d}V = (1{+} C_{diss}) \left(\int_V \frac{1}{2} \rho u^2 \mathrm{d}V {+} \int_V \frac{1}{2} b^2 \mathrm{d}V \right).
\end{equation}

From the self-similarity arguments, we expect that the TKE and TME are proportional to each other (i.e., TME ${\propto}$ TKE). Let us consider, $\int_V \frac{1}{2} b^2 \mathrm{d}V = C_{ep} \int_V \frac{1}{2} \rho u^2 \mathrm{d}V$. Under this assumption, the above equation can be rewritten as
\begin{equation}
    - \int_V \delta \rho g x_3 \mathrm{d}V = (1{+} C_{diss})(1{+}C_{ep}) \left(\int_V \frac{1}{2} \rho u^2 \mathrm{d}V \right).
    \label{hi}
\end{equation}

The TKE constitutes energy from each component of velocity (i.e., $\int_V \frac{1}{2} \rho u^2 \mathrm{d}V = \sum_{i=1}^{3} \int_V \frac{1}{2} \rho u_i^2 \mathrm{d}V$). The self-similarity implies that the TKE along the homogeneous directions ($x_1$, $x_2$) and the non-homogeneous direction ($x_3$) are proportional. This was indirectly proposed and evidenced by \cite{Youngs1991, Linden_Redondo_Youngs_1994} for the HD RTI. \cite{Youngs1991} proposed that in the self-similar regime, the ratio of $\frac{1}{2} \rho u_i^2, i = [1, 3]$ and GPE is constant. Consequently, the ratio of horizontal and vertical components of TKE should also be constant. That is, $\sum_{i=1}^{2} \int_V \frac{1}{2} \rho u_i^2 \mathrm{d}V {\propto} \int_V \frac{1}{2} \rho u_3^2 \mathrm{d}V$. Considering $C_{aniso}$ as the proportionality constant, equation \ref{hi} simplifies to

\begin{equation}
    - \int_V \delta \rho g x_3 \mathrm{d}V = (1{+} C_{diss})(1{+}C_{ep}) (1{+}C_{aniso}) \left(\int_V \frac{1}{2} \rho u_3^2 \mathrm{d}V \right).
\end{equation}

To derive the equation of mixing layer height ($h$) in terms of $t$, we need to write the quantities on the left and right-hand sides of the above equation in terms of $h$ or $\partial_t h$. In the region outside the mixing layer, $\delta \rho {=} 0$. Hence, the volumetric integral of $\delta \rho x_3$ is proportional to $V_m \Delta \rho h$, where $V_m$ is the volume of the mixing layer $(V_m {=} L_x L_y h)$ and $\Delta \rho$ is the density difference. Let,
\begin{equation}
    \int_{V} \delta \rho x_{3} \mathrm{d}V = C_{com} L_x L_y h^2 \Delta \rho. 
    \label{C5B}
\end{equation} 
The velocity of the mixing layer front depends on the growth rate of the mixing layer height, the characteristic length scale of the MRTI. Therefore $u_3 \propto \partial_t h$. In the region outside the mixing layer, $u_3 {\approx} 0$. Hence, $\int_V \frac{1}{2} \rho u_3^2 \mathrm{d} V {\propto} V_m \overline{\rho} (\partial_t h)^2$, where $\Bar{\rho}$ is the volume averaged density. Let
\begin{equation}
    \int_V \frac{1}{2} \rho u_3^2 \mathrm{d}V = \frac{1}{C_{gr}}  L_x L_y h \overline{\rho} (\partial_t h)^2.
    \label{C4B}
\end{equation}
We chose the constant as $1/C_{gr}$ for the ease of calculating $C_{gr}$ later when the RTI is modelled numerically. Implementing the above scaling, we get,
\begin{equation}
    C_{com} g L_x L_y \Delta \rho h^2 = \frac{(1{+}C_{diss})(1{+}C_{ep}) (1{+}C_{aniso})}{C_{gr}} h (\partial_t h)^2 L_x L_y \Bar{\rho},
\end{equation}

Integrating the above equation with time using the separation of variables method, we get
\begin{equation}
    h = \alpha_{mhd} \mathcal{A}g t^2 {+} 2 \sqrt{\alpha_{mhd} \mathcal{A} g h_0} t {+} h_0, \text{ where } \alpha_{mhd} (\mathcal{A}, B_0) = \frac{ C_{com} C_{gr}}{2 (1{+}C_{diss})(1{+}C_{ep}) (1{+}C_{aniso})}, 
    \label{intermediate}
\end{equation}
where $h_0$ is the height of the mixing layer at $t {=} 0$ assuming the mixing layer height had a quadratic variation from $t {=} 0.$ Equation \ref{intermediate} confirms that the mixing layer height of non-linear MRTI grows quadratically in time even in the presence of magnetic field, similar to the HD RTI. While the magnetic field do not appear explicitly in the `$h$' equation, the coefficients $C_{diss}$, $C_{ep}$, $C_{aniso}$, $C_{gr}$, $C_{com}$, and consequently $\alpha_{mhd}$ vary with the magnetic field strength and the Atwood number. Thus, the magnetic field and Atwood number influence the non-linear growth of instability.

From the above equation, we see that the non-linear growth of instability is controlled by:  \vspace{-5pt}
\begin{enumerate}[i)]
    \item the ratio of total energy dissipated to the total turbulent energy ($C_{diss}$), \vspace{-5pt}
    \item the energy partition between TME and TKE ($C_{ep}$), \vspace{-5pt}
    \item distribution of TKE among homogeneous and non-homogeneous components ($C_{aniso}$), \vspace{-5pt}
    \item the scaling of the non-homogeneous component of TKE and GPE with their respective non-dimensional forms ($C_{gr}$, $C_{com}$). \vspace{-5pt}
\end{enumerate}
Analytical estimation of these coefficients is not possible, necessitating the numerical modelling of MRTI to determine $\alpha_{mhd}$. However, the advantage of having a formula for $\alpha_{mhd}$ is the understanding it brings as to what parameters play an important role in the non-linear growth of MRTI. Further, using the above formula and a systematic study of MRTI at different magnetic field strengths and Atwood numbers, we can understand how the magnetic field strength and the Atwood number influence the non-linear growth. 

Having determined the temporal variation of mixing layer height, we can predict the temporal variation of other quantities like momentum, turbulent magnetic field, TKE, and TME. We know that the (vertical) velocity of the mixing layer is of the order $\partial_t h$ (i.e., $O(t)$). The volume-averaged momentum is expected to be of order $O(h) O(t)$, i.e., $O(t^3)$. Volume averaged TKE is of the order $h (\partial_t h)^2$ i.e., $O(t^2) O(t^2) = O(t^4)$. In terms of mixing layer height, $\int_V \mathrm{TKE}\mathrm{d}V$ is of the order $h^2$. From self-similar scaling,  $\int_V \mathrm{TME}\mathrm{d}V$ is expected to be of the same order, $O(t^4)$ or $O(h^2)$. The $t^4$ variation of TKE, TME agrees with the temporal scaling of TKE, TME reported in \citet{Stone2007a}. The $h^2$ scaling of TKE and GPE was proposed in HD RTI studies (a limit case of MRTI with $B_0 = 0$) by \cite{Youngs1991, Linden_Redondo_Youngs_1994}.

\subsection{Summary}
To summarize $\S$\ref{h-derivation}, an equation for the mixing layer height `$h$' in the self-similar regime is derived. The height of the mixing layer was proved to grow quadratically in time (equation \ref{intermediate}), similar to the HD case. A formula for the non-linear growth constant ($\alpha_{mhd}$) was deduced (equation \ref{intermediate}). The formula elucidated various factors that could play a crucial role in the non-linear growth of instability. However, the determination of the $\alpha_{mhd}$ demands determining the coefficients, which necessitates the numerical modelling of MRTI. 

Towards this end, we perform numerical simulations. The simulations will also validate the conclusions of the analytical theory. While the coefficients change with magnetic field and Atwood number, we will focus on the former in this paper. To understand the influence of magnetic field on the $\alpha_{mhd}$ we simulate MRTI at different magnetic field strengths and study the variation of these coefficients and $\alpha_{mhd}$.

\section{Numerical methodology} \label{Methodology}

While a formula for $\alpha_{mhd}$ was derived, it's value remains unknown unless the values of $C_{diss}$, $C_{ep}$, $C_{aniso}$, $C_{gr}$, $C_{com}$ are known. Determining these coefficients demands performing numerical simulations of non-linear MRTI. Further, the derivations of the mixing layer height are based on certain assumptions. To validate the above assumptions and corroborate the conclusions of the analytical study, we need to simulate the MRTI numerically. This section details the methodology of the numerical simulations.

In the present study, MRTI is modelled by superimposing a high density fluid $(\rho_h)$ over a low density fluid $(\rho_l)$ in the presence of uniform, unidirectional magnetic $(\mathbf{B_0})$ and gravitational $(\mathbf{g})$ fields as shown in figure \ref{MRTI_config}(a). The magnetic and gravitational fields are along the $x_1$ and $x_3$ directions, respectively. The numerical modelling is performed using Dedalus \citep{Dedalus2020}, an open-source, parallelized computational framework to solve the partial differential equations using the pseudo-spectral method. 

The divergence of velocity is of the order $\frac{1}{Re Sc} \nabla \cdot \left(\frac{1}{\rho} \nabla\rho \right)$. The influence of velocity divergence is prominent only at length scales where the density gradient is at least of the order of $ReSc$. In the limit of moderate density gradients and large $Re Sc$, the term $\mathbf{\nabla} {\cdot}  \mathbf{u}$ becomes negligible, as seen in the numerical HD RTI study by \cite{Cabot_2013}. This is typically the case in most astrophysical systems where the order of magnitude of density gradients is much smaller than the $Re Sc$ i.e., $O(Re Sc) \gg O \left(\nabla \cdot \left( \frac{1}{\rho} \nabla\rho \right) \right)$ making $\mathbf{\nabla} {\cdot}  \mathbf{u} \approx 0$. Hence, in the present study, we solve an \textit{approximation of the Variable-Density} \citep{Soulard2012, Zhou_2024book, LIVESCU_RISTORCELLI_2007} non-ideal MHD governing equations:
\begin{subequations}
    \begin{align}
        \partial_t \mathbf{u} - \frac{1}{Re} \mathbf{\nabla}^2 \mathbf{u} & = {-} (\mathbf{u} {\cdot} \mathbf{\nabla}) \mathbf{u} {-} \frac{1}{\rho} \mathbf{\nabla} p {-} \frac{\delta \rho}{\rho}\mathbf{g} {-} \frac{1}{\rho} (\mathbf{B} {\cdot}  \mathbf{\nabla}) \mathbf{B}, \label{MHDeqnNS} \\
        \partial_t \rho - \frac{1}{Re Sc} \mathbf{\nabla}^2 \rho & = {-} (\mathbf{u} {\cdot} \mathbf{\nabla}) \rho,  \\
        \mathbf{\nabla} {\cdot}  \mathbf{u} & = 0, \\
        \partial_t \mathbf{B} - \frac{1}{Re Pr} \mathbf{\nabla}^2 \mathbf{B} {-} c_p^2 \mathbf{\nabla}(\mathbf{\nabla} {\cdot} \mathbf{B}) & = (\mathbf{B} {\cdot} \mathbf{\nabla}) \mathbf{u} {-} (\mathbf{u} {\cdot} \mathbf{\nabla}) \mathbf{B}.
    \end{align}
    \label{MHDeqns_1}
\end{subequations}
The fluctuating pressure ($p$) includes both fluid and magnetic pressures. The Prandtl number ($Pr$) is defined as the ratio of fluid viscosity to magnetic diffusivity ($\mathcal{\eta}$) (i.e., $Pr = \nu/\mathcal{\eta}$). The values of $Re, Pr,$ and $ Sc$ are set to $10^{4}, 1,$ and $1$ respectively. The solenoidal condition for the magnetic field is ensured through the divergence cleaning term ($c_p^2 \mathbf{\nabla}(\mathbf{\nabla}.\mathbf{B})$) \citep{DEDNER_2002}. The value of $c_p$ is set to 10 so that the solenoid condition is satisfied to machine precision at all times. A simple Laplacian density diffusion is used in equation \ref{MHDeqns_1}b to ensure numerical stability. As the interface elongates, the interface thins naturally and becomes unresolved if this term is not included.

The pressure is solved by splitting the pressure term $\left( \frac{1}{\rho} \nabla p \right)$ as a linear part $(\nabla p')$ and a nonlinear part $\left( \frac{p'}{\rho} \nabla \rho \right)$, where $p' = p/\rho$. The linear pressure term is solved from the implicit time stepping scheme, and the non-linear component of pressure is updated by explicit time stepping. The linear pressure term is determined based on the divergence-free velocity constraint. The sparse matrices are solved using SuperLU and UMFPACK libraries, which use the LU Factorisation method. The time step is determined from the CFL condition with a factor of safety of 0.2. The time stepping scheme used is a four-stage third-order Runge-Kutta \citep{ASCHER1997}.

Here, MRTI is studied in 3D with periodic boundary conditions in all directions. A domain of length $L_1 {\times} L_2 {\times} L_3 {=}$ $ 4 {\times} 4 {\times} 6$ units $\left( x_1:[0, L_1], x_2:[0, L_2], x_3: \left[ -\frac{L_3}{2}, \frac{L_3}{2} \right] \right)$ with a resolution of $512 {\times} 512 {\times} 768$ is taken. The acceleration due to gravity is taken as 1. Since the purpose of the study is to understand the role of the magnetic field on self-similarity, we present the results for a single Atwood number. We choose to study the problem in the largest density ratio possible. Due to the choice of using a spectral solver, the maximum density ratio that can be used is limited. This is due to the ringing artifact caused by sharp density jumps. While density diffusion aids in smoothing the density jumps at the grid scale, large density ratios lead to sharp density gradients, resulting in ringing. For the chosen density diffusion coefficient, test simulations with increasing Atwood number showed the ringing effect for $A {>} 0.5$. Hence, the Atwood number of 0.5 with $\rho_h {=} 3$ and $\rho_l {=} 1$ was chosen. 

The initial density profile is given by equation \ref{rho_profile} and shown in figure \ref{MRTI_config}(a). 
\begin{equation}
    \rho = 1 {-} \frac{(\rho_h {-} \rho_l)}{2} \left[ \tanh{\left(\frac{x_3 {-} 0.45 L_3}{0.05} \right)} {+} 1 \right] {+} \frac{(\rho_h {-} \rho_l)}{2} \left[ \tanh{\left(\frac{x_3}{0.05} \right)} {+} 1 \right] 
    \label{rho_profile}
\end{equation}
The above density profile is chosen to facilitate periodicity in the $x_3$-direction. The profile results in two interfaces, one at $x_3 {=} 0$ and the other at $x_3 {=} 0.45 L_3$, as shown in Figure \ref{MRTI_config} (a). The transition between the two densities is made continuous using a hyperbolic tangent profile with a half width of $l {=} 0.05$ (see equation \ref{rho_profile}). Thus, there are $\approx 7$ grid points throughout the width of the interface. As the MRTI evolves, the instability elongates the interface of the two fluids over time. As the bubbles (and spikes) evolve, the field lines get bundled around the bubbles (and spikes), resulting in a thinner interface over time. While the density diffusion aids in smoothing the density profile, it is important to ensure that the interface is initially resolved over an adequate number of grid points, so that there are a sufficient number of grid points at any point in time in the evolution. Hence, we chose a half-width of 0.05 so that the interface remains adequately resolved. 

The upper interface is left unperturbed. Even if perturbed, it does not undergo RTI due to its stable configuration (high-density fluid supporting the low-density fluid). An advantage of the stable interface close to the top boundary is that it acts as a marker of boundary influence on flow structures. In the current system, as long as the top interface remains unaffected by the rising plumes, we can consider that the boundary influences are absent. The two-interface density structure is common among experimental \citep{Suchandra_Ranjan_2023, DALZIEL2021} and numerical \citep{Briard_Gréa_Nguyen_2024, Briard_2022} studies. 

The lower interface $(x_3 {=} 0)$ of the system is perturbed by a vertical velocity. The perturbation is of the form
\begin{equation}
    u_3 {=} \left(\sum^{64}_{k_i=0} a_i \sin \left( \frac{2 \pi k_i x_1}{L_1} {+} \phi_i \right) \times \sum^{64}_{k_j=0} a_j \sin \left( \frac{2 \pi k_j x_2}{L_2} {+} \phi_j \right)  - (\underbrace{a_0 \sin \phi_0}_\text{$i=0$} \times \underbrace{a_0 \sin \phi_0}_\text{$j=0$} ) \right) e^{{-}(x_3^2/0.01)}.
    \label{white_noise_3d}
\end{equation}
Here, $a$, $k$, and $\phi$ are chosen between [-0.001, 0.001], [1, 64], and [0, $\pi$], respectively. We introduce a wide range of wave modes as suggested by \citet{ramaprabhu_2005, dalziel_1999, Dimonte2004, GLIMM2001}. The perturbations decay in a Gaussian profile about the interface $(x_3 {=} 0)$. The system has undular $(k_i \neq 0, k_j = 0)$, interchange $(k_i = 0, k_j \neq 0)$, and a wide range of modes in between $(k_i \neq 0, k_j \neq 0)$.

To test the self-similar evolution of MRTI and further ensure its validity across different magnetic field strengths, we run MRTI simulations over a range of magnetic field strengths between $1\%B_c$ and $25\% B_c$ (for $B_c$, see equation \ref{Bc}). 
Note that, $B_c$ is originally defined for a single-mode undular perturbation. For consistency, we choose the $B_c$ based on the smallest undular mode i.e., $\mathbf{k} = (k_x, k_y) = (1, 0)$, since $k_x$ is parallel to the imposed magnetic field. Typically, $B_c$ is used to non-dimensionalize the imposed magnetic field strength \citep{Stone2007a, Jun1995}, even though $B_0 = B_c$ may not suppress the instability completely due to the presence of interchange modes. Following the previous studies, we non-dimensionalize the imposed magnetic field strength with $B_c$, as a way to characterize the imposed magnetic field strength.
We ensure that the vertical confinement of the mixing layer is avoided at all times. The volume fraction of the mixing layer is always ensured to be less than $50\%$ of the available mixing region. This can be seen from figure \ref{MRTI_config}(d). 

\section{Results from numerical experiments} \label{Results}

We have made several theoretical predictions about the self-similarity of MRTI. In $\S$\ref{self-similarity} we reported that the influence of the initial magnetic field term decays as $1/t$ relative to other non-linear dynamics terms (cf. equations \ref{self_similar_TKE}, \ref{self_similar_TME}). Following this, we deduced that MRTI approaches towards HD self-similar scaling at late time. Based on this, in $\S$\ref{h-derivation}, we derived equations of mixing layer height ($h$) and the non-linear growth constant ($\alpha_{mhd}$) in the self-similar regime. The derived `$h$' equation has quadratic variation in time. We proposed that volume-averaged TKE and TME scale as $h^2$. The expression of $\alpha_{mhd}$ showed that it is a function of five \textit{undetermined} coefficients --- $C_{diss}$, $C_{ep}$, $C_{aniso}$, $C_{gr}$, and $C_{com}$. However, the `$h$' equation and the derivation of $\alpha_{mhd}$ expression are based on a few assumptions, which remain \textit{unverified}. Towards corroborating the analytic conclusions, verifying the assumptions, and determining the coefficients, we perform numerical simulations. These results will be discussed in this section. In $\S$\ref{self-similarity_validation}, we verify the temporal variation of the initial magnetic field term relative to other non-linear terms. In $\S$\ref{sec:estimation}, we verify the quadratic growth of mixing layer height and determine the coefficients and $\alpha_{mhd}$. In $\S$\ref{Mag_field_effect}, we study the role of the magnetic field on the coefficients and $\alpha_{mhd}$.

\subsection{Self-similarity and non-linear growth of MRTI}

\subsubsection{Decay of imposed magnetic field term} \label{self-similarity_validation}
\begin{figure}
     \centering
     \begin{subfigure}[b]{0.49\textwidth}
         \centering
         \includegraphics[width=\textwidth]{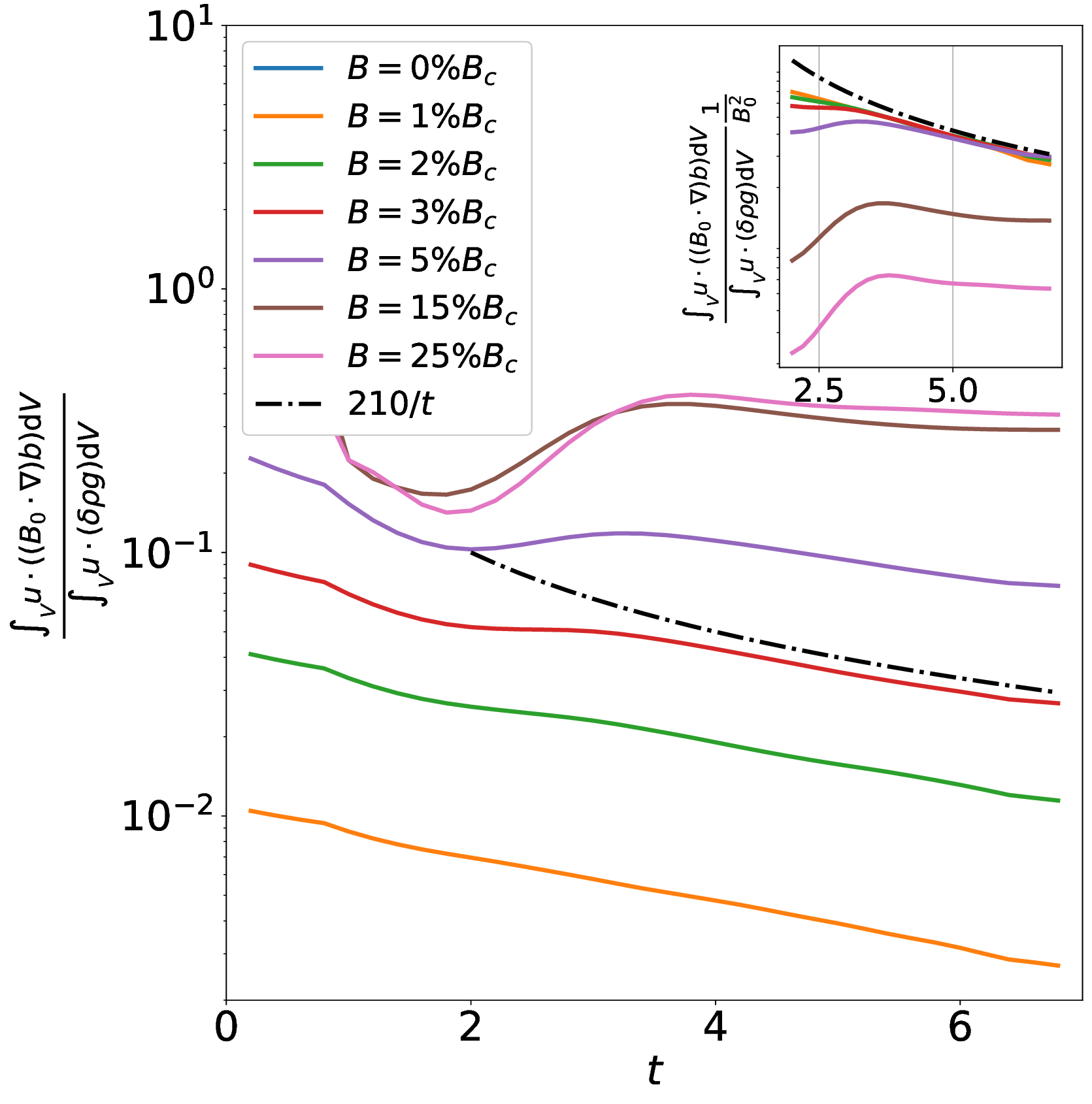}
     \end{subfigure}
     \hfill
     \begin{subfigure}[b]{0.49\textwidth}
         \centering
         \includegraphics[width=\textwidth]{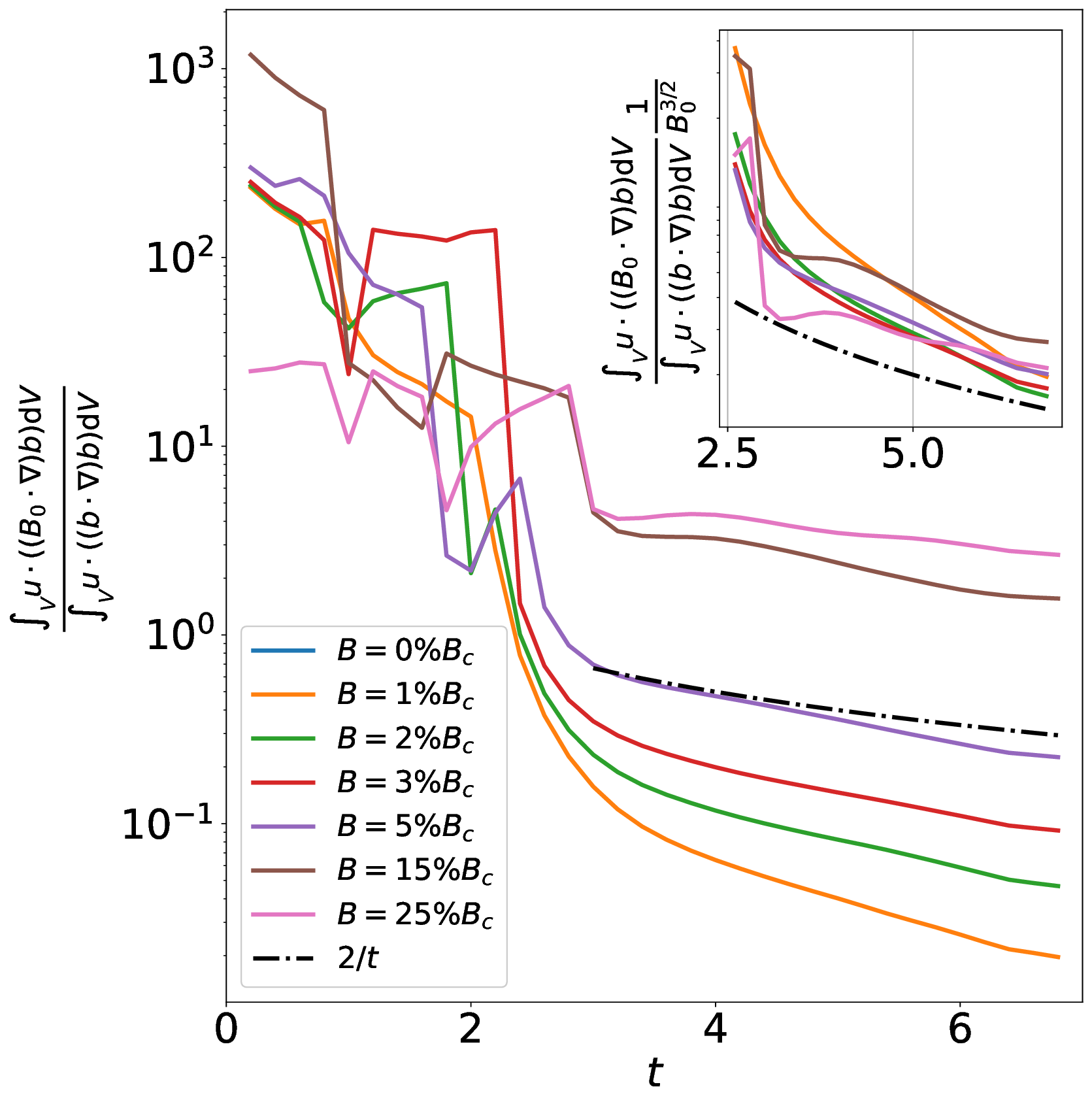}
     \end{subfigure}
     \caption{Temporal variation of the initial magnetic field term $\left( \int_V u \cdot (B_{0} \cdot \nabla) b \mathrm{d}V \right)$ relative to: \textit{(left)} the gravity term $\left( \int_V (\delta \rho g \updelta_{i3}) \cdot u \mathrm{d}V \right)$ and \textit{(right)} the Lorentz force term $\left( \int_V u_i (b \cdot \nabla) b_i \mathrm{d}V \right)$ of TKE equation \ref{TKE_eqn} for different magnetic field strengths. In the inset figures, we show the quantities on the Y-axis scaled with $B_0^2$ (left) and $B_0^{3/2}$ (right) so that the validity of the scaling law for different magnetic field cases is lucid.}
     \label{B0udb_rhov}
\end{figure}

The MRTI may not be always self similar, but we expect it approaches towards self-similarity as the influence of initial magnetic field term decays with time as $1/t$ relative to the other terms of TKE like the GPE $\left( \int_V \delta \rho g_i u_i \updelta_{i3} \mathrm{d}V \right)$, Lorentz force term $\left( \int_V u_i (b \cdot \nabla) b_i \mathrm{d}V \right)$ (cf. $\S$\ref{self-similarity}). This is necessary for the convergence of MRTI towards HD self-similarity. To validate this, we plot the ratio of initial magnetic field term $\left( \int_V u_i (B_{0} \cdot \nabla) b_i \mathrm{d}V \right)$ to the GPE term (figure \ref{B0udb_rhov}(left)) and the Lorentz force term (figure \ref{B0udb_rhov}(right)) over time. As predicted by the analytical theory, the strength of the imposed magnetic field decreases as $1/t$ relative to the GPE and Lorentz force term for all magnetic field strengths. A more convincing demonstration of the $1/t$ scaling of these quantities is shown in the inset figures, where we plot the quantities scaled with imposed magnetic field strength ($B_0$) appropriately. Thus, with diminishing influence of magnetic field, the mixing layer approaches towards self-similarity, and the growth tends towards the quadratic HD scaling. 

The quadratic growth of mixing layer was explicitly derived in $\S$\ref{h-derivation}. The volume-averaged TKE and TME are proposed to have $h^2$ variation. To confirm these numerically, we plot the temporal variation of $h/t^2$ for different magnetic field strengths in figure \ref{h,tke,tme}(left) (solid lines). Typically, the \textit{height of the mixing layer} is calculated based on the mixing parameter, $\Theta (x_3) {=} 4 \left \langle \left( \frac{ \rho  {-} \rho_l}{\rho_h {-} \rho_l} \right) \left( \frac{\rho_h {-} \langle \rho \rangle}{\rho_h {-} \rho_l} \right) \right \rangle$ \citep{Stone2007a}, where $x_3$ is the direction of mixing layer growth. $\langle \star \rangle$ refers to averaging along the statistical homogeneous directions ($x_1, x_2$). Outside the mixing layer, where the fluid density is either $\rho_h$ or $\rho_l$, $\Theta {=} 0$. In the mixing layer, $\Theta$ lies between $(0, 1]$. $\Theta {=} 1$ is the well mixed condition when $\langle \rho \rangle {=} \frac{\rho_h {+} \rho_l}{2}$. The boundaries of the mixing layer are defined by choosing a threshold value of $\Theta$. The boundaries of the mixing layer above and below the center line ($x_3 = 0$) represent the height of the bubble ($h_b$) and height of the spike ($h_s$), respectively. The height of the mixing layer ($h$) is the sum of the magnitudes of $h_b$ and $h_s$. This is the method we use in this study for consistency with other MRTI studies like \citet{Stone2007a, Jun1995, Carlyle2017}. However, other ways of estimating mixing layer height are detailed in \citet{baltzer_livescu_2020}. In the regime of quadratic growth, $h/t^2$ is expected to be approximately constant. However, the figure \ref{h,tke,tme}(left) shows that for field strengths $B_0 \leq 5\%B_c$, the $h/t^2$ curve slopes down, denoting that the quadratic growth could be an overestimate. Hence, we also plot the temporal variation of $h/t$ (dashed line in figure \ref{h,tke,tme}(left)), which appears to be flatter compared to $h/t^2$ lines, hinting that the mixing layer growth is more linear than quadratic. Note that this does not prove that the quadratic growth (equation \ref{intermediate}) is invalid. It is possible that the coefficient of linear growth ($2 \sqrt{\alpha_{mhd} \mathcal{A} g h_0} $) is dominant compared to quadratic growth ($\alpha_{mhd} \mathcal{A} g$). To test this conjecture, we calculate these quantities in $\S$\ref{sec:estimation}. For the stronger magnetic fields $B_0 \geq 15\%B_c$, the mixing layer growth becomes near quadratic. Similarly, the $t^4$ scaling of the volume-averaged TKE and TME was found to be an overestimate. However, the volume averaged TKE (plotted as solid lines) and TME (plotted as dashed lines) scale as $h^2$ (see figure \ref{h,tke,tme}(right)).

\begin{figure}
     \centering
     \begin{subfigure}[b]{0.49\textwidth}
         \centering
         \includegraphics[width=\textwidth]{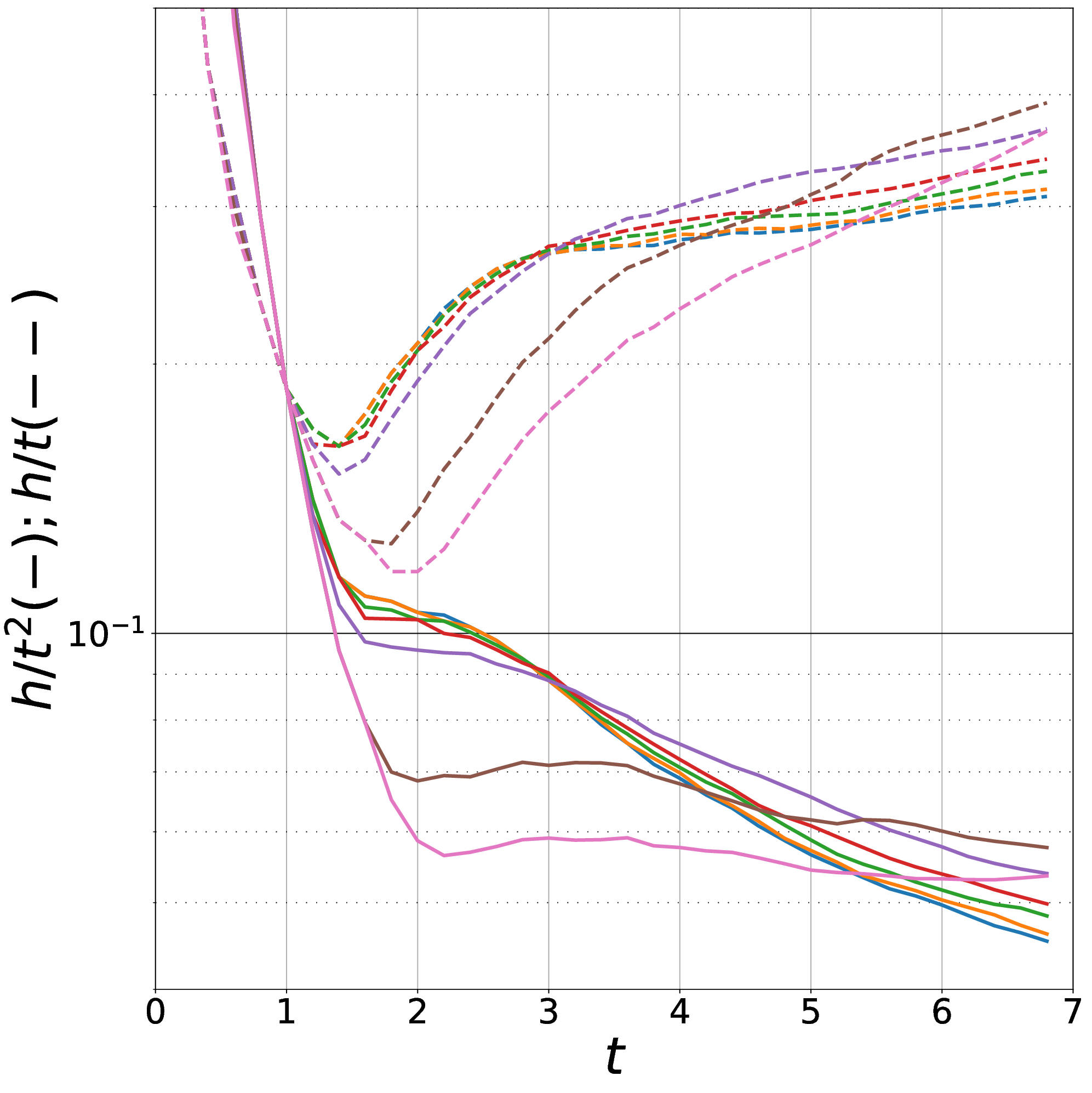}
     \end{subfigure}
     \hfill
     \begin{subfigure}[b]{0.49\textwidth}
         \centering
         \includegraphics[width=\textwidth]{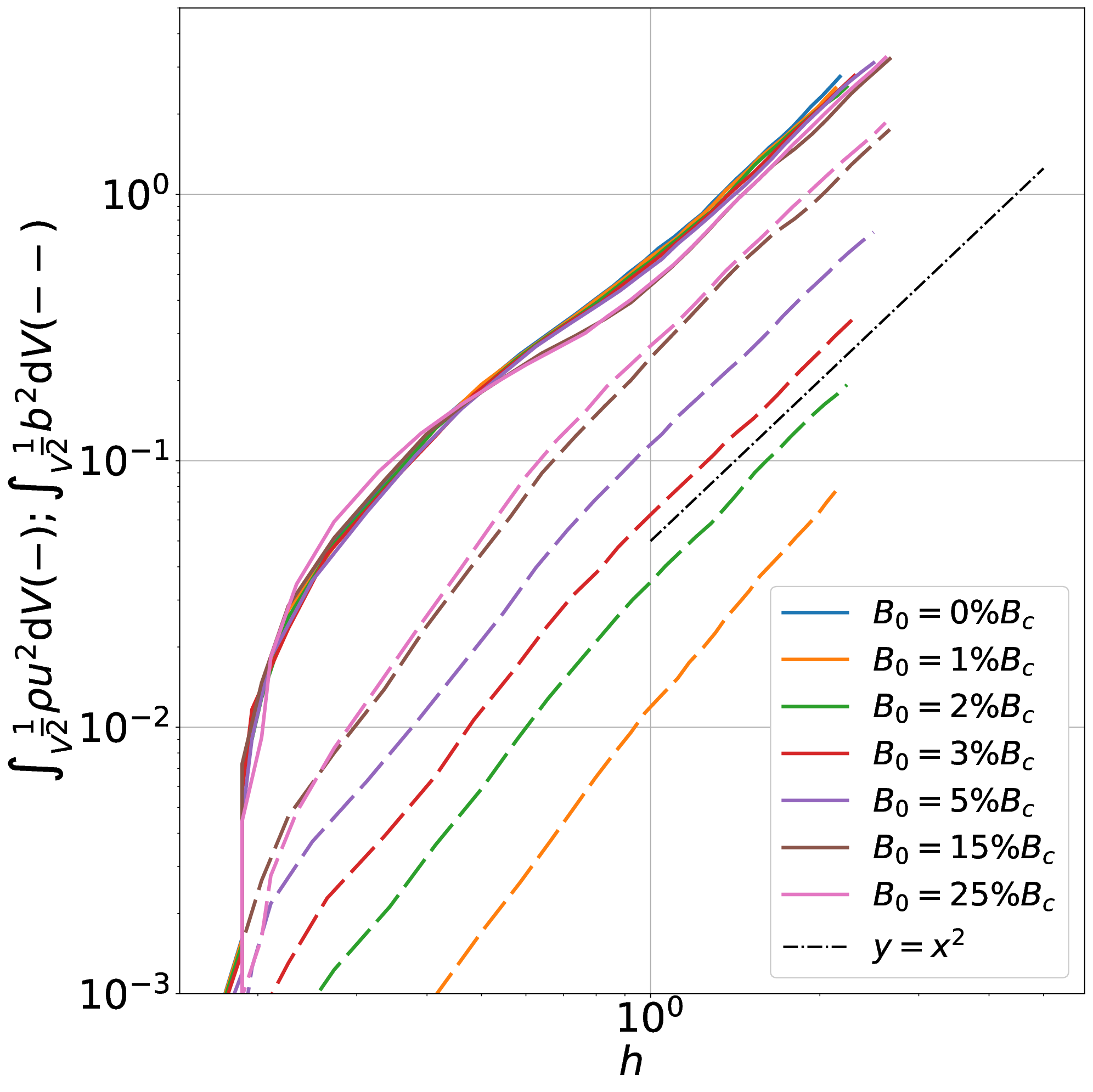}
     \end{subfigure}
     \caption{\textit{(left)} Temporal variation of the mixing layer height ($h$) scaled with $t^2 (-)$ and $t (- -)$; \textit{(right)} Scaling of turbulent magnetic $(- -)$ and kinetic $(-)$ energies with the mixing layer height. The legend is the same for both sub-figures.}
     \label{h,tke,tme}
\end{figure}

\subsubsection{Estimation of non-linear growth constant} \label{sec:estimation}

Towards comparing the quadratic and linear growths, we determine the $t^2$ and $t$ coefficients using the curve-fitting technique. Consider a quadratic function $\hat{y} {=} a \hat{x}^2 {+} 2 \sqrt{a b} \hat{x} {+} b$, where the input parameters, $\hat{x}$ and $\hat{y}$, are $t$ and $h$, respectively. Since the self-similarity is achieved at $t \approx 4$ in these simulations, we consider the data from $t \geq 4$. Therefore, $\Hat{x} = t_{t=4} \text{ to } t_{t=7}$, and $\Hat{y} = h_{t=4} \text{ to } h_{t=7}$. $a$ and $b$ represent the $\alpha_{mhd} \mathcal{A} g$ and $h_0$ (cf. equation \ref{intermediate}). The above quadratic function is chosen to restrict the number of free variables in the quadratic equation to the number of input parameters. Further, this also incorporates the dependency between the coefficients. 

\begin{figure}
     \centering
     \begin{subfigure}[b]{0.50\textwidth}
         \centering
         \includegraphics[width=\textwidth]{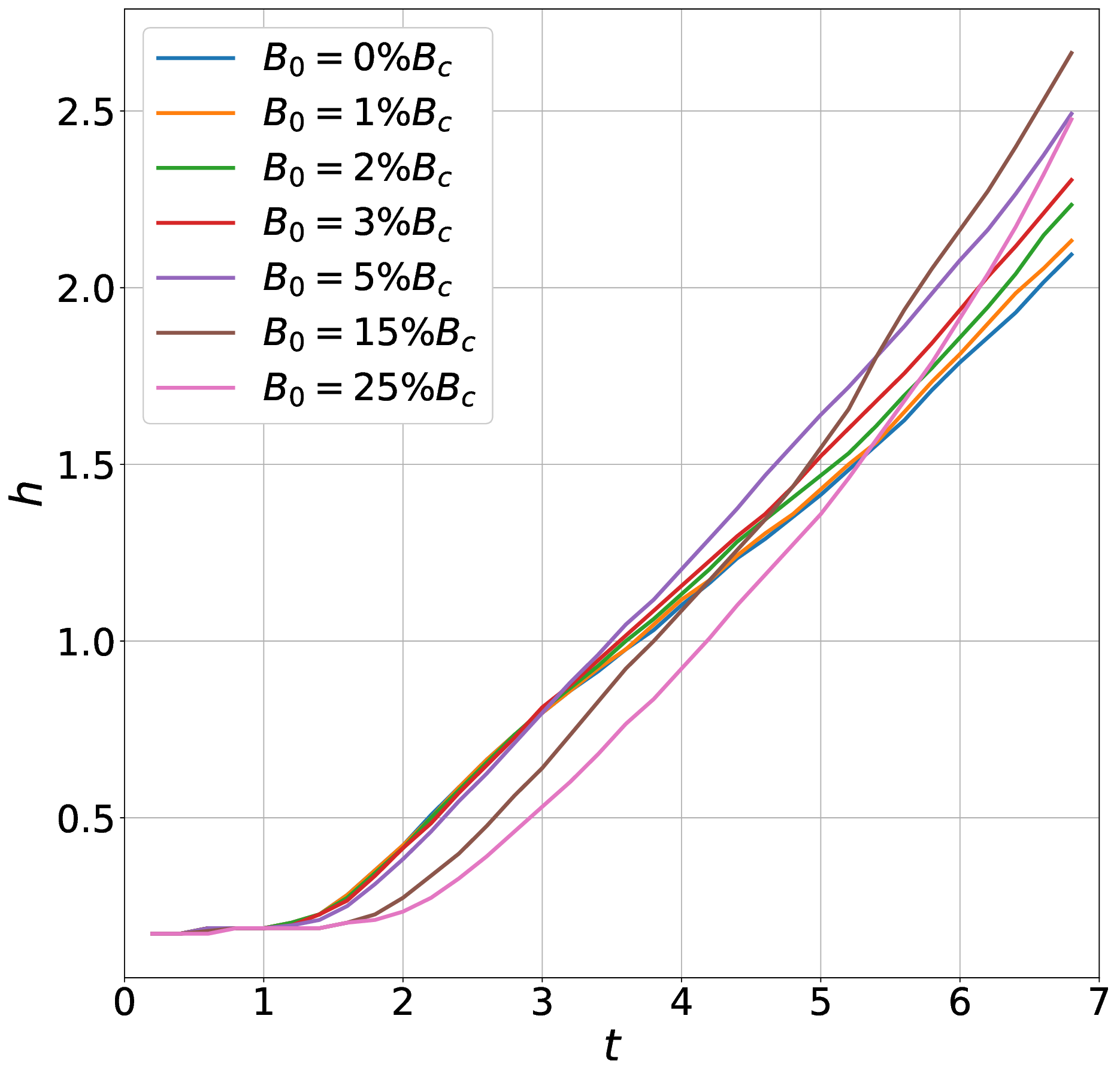}
     \end{subfigure}
     \hfill
     \begin{subfigure}[b]{0.49\textwidth}
         \centering
         \includegraphics[width=\textwidth]{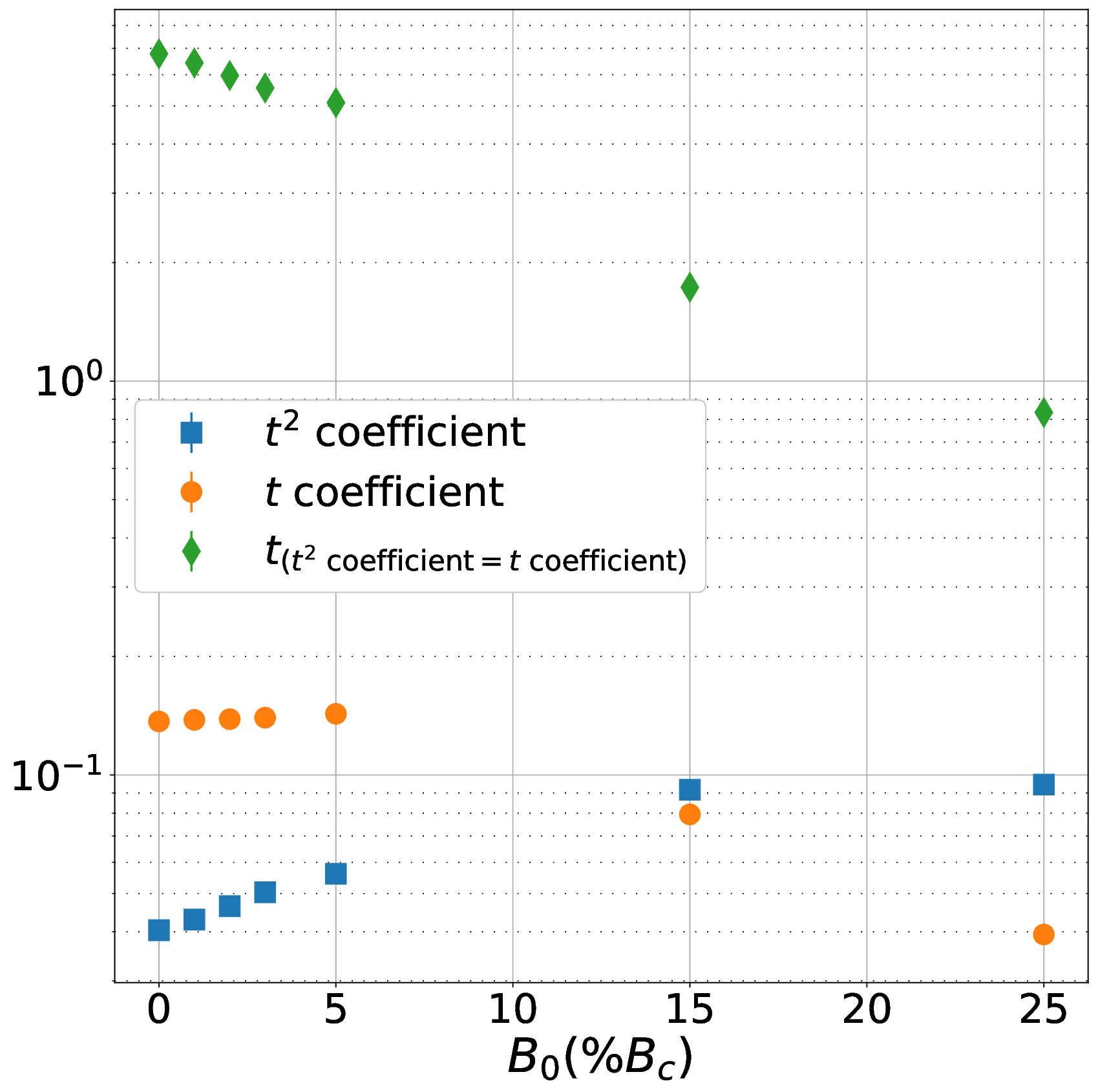}
     \end{subfigure}
     \caption{\textit{(left)} Temporal variation of $h$ for different magnetic field strengths. \textit{(right)} Coefficients of $t^2$ ($\textcolor{blue}{\blacksquare}$) and $t$ ($\textcolor{orange}{\CIRCLE}$) of mixing layer height equation (equation \ref{intermediate}) for different magnetic field strengths. The data to estimate $\alpha_{mhd}$ is sampled from $t \geq 4$. $\textcolor{ForestGreen}{\blacklozenge}$ represents the time instant beyond which the quadratic growth dominates the linear growth of the mixing layer.}
     \label{h_t}
\end{figure}

Figure \ref{h_t}(left) shows the temporal variation of $h$. Figure \ref{h_t}(right) shows the coefficients of quadratic and linear growth. The small value of $\alpha_{mhd}$ makes the linear growth prominent, since $\sqrt{\alpha_{mhd}} {\gg} \alpha_{mhd}$, $\forall$ $\alpha_{mhd} {\ll} 1$. Thus, the mixing layer growth is predominantly linear in the weak magnetic fields at the early times. This behavior was also observed in the experimental studies of HD RTI, where the Atwood number is the same as the current study \citep{Suchandra_Ranjan_2023}. To approximate the time at which the quadratic growth becomes dominant, we calculate the time at which the quadratic and the linear coefficient of equation \ref{intermediate} are equal. This time instant is marked as green diamonds in figure \ref{h_t}(right). Note that the reference time here is not $t = 0$, but $t=4$. As the strength of the magnetic field increases, the quadratic growth coefficients become stronger and hence the quadratic growth behaviour onsets earlier.

\subsubsection{Verification of self-similarity assumptions} \label{section:verification}

Towards deriving the equation of mixing layer height in $\S$\ref{h-derivation}, several assumptions were made, which lead to different coefficients that play a key role in $\alpha_{mhd}$. The self-similarity of these coefficients remains unverified. This section aims to validate these assumptions for different magnetic field strengths.
\begin{figure}
    \centering
    \begin{subfigure}[b]{0.48\textwidth}
         \centering
         \includegraphics[width=\textwidth]{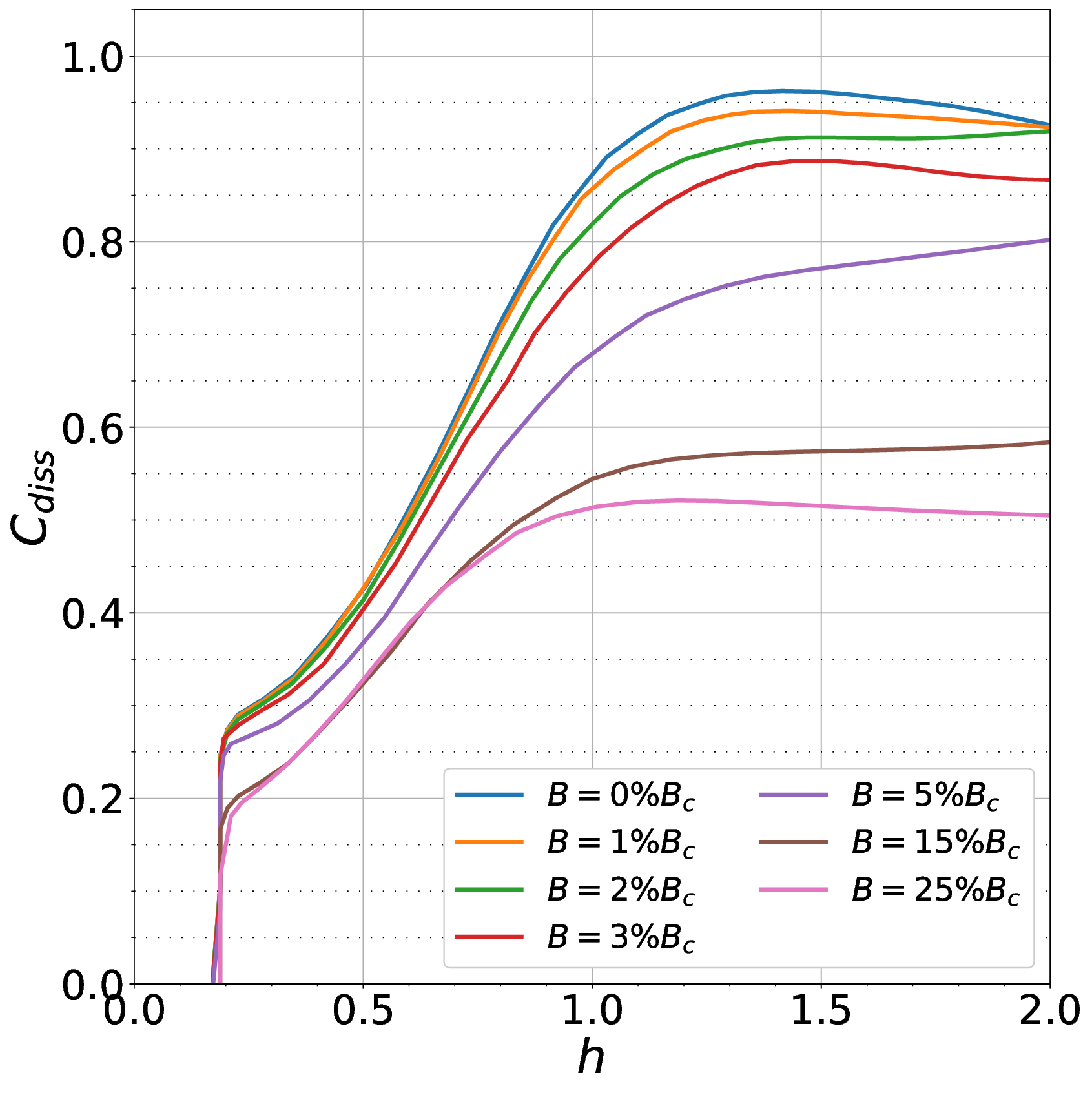}
     \end{subfigure}
     \hfill
     \begin{subfigure}[b]{0.50\textwidth}
         \centering
         \includegraphics[width=\textwidth]{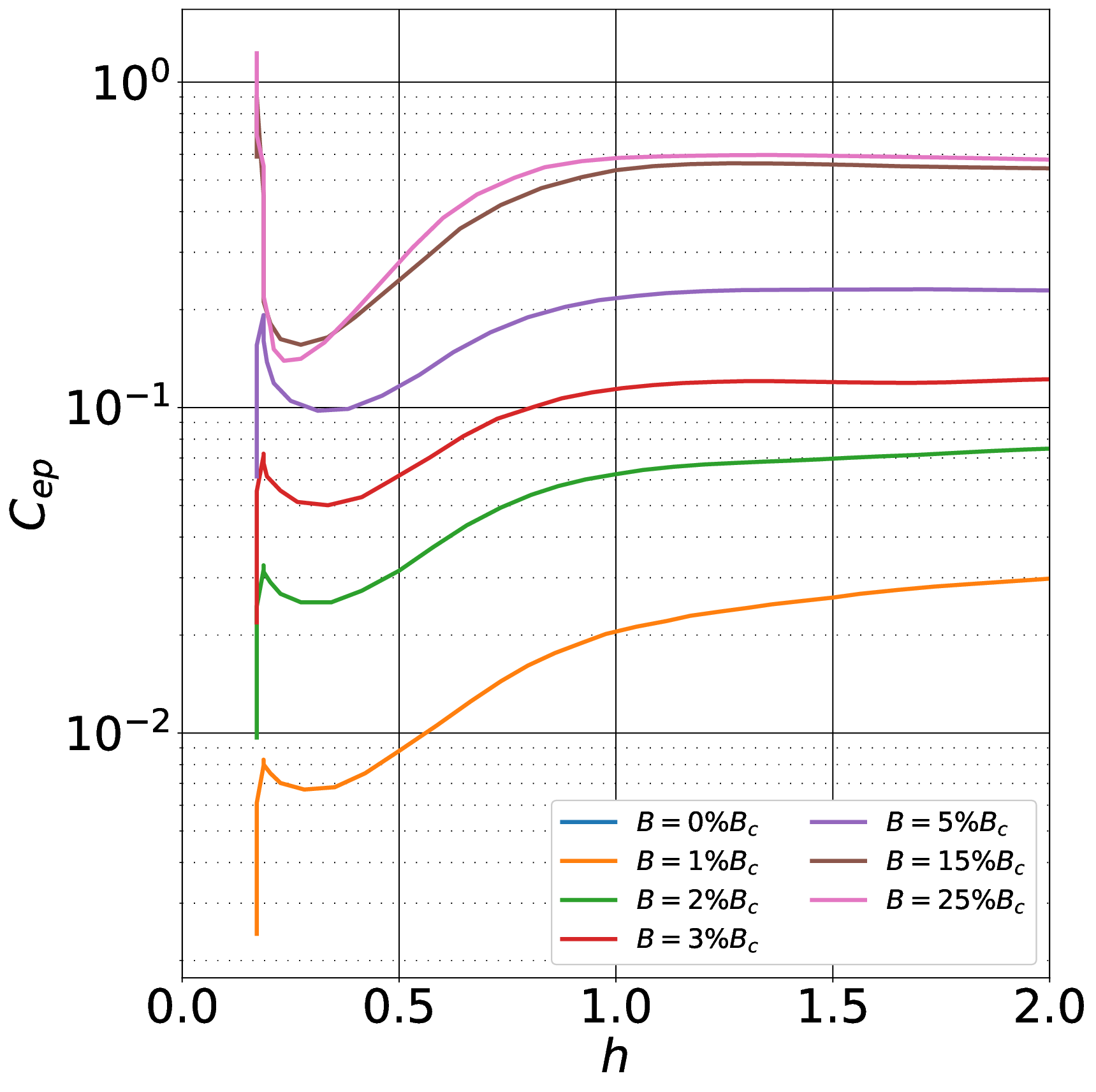}
     \end{subfigure}
    \caption{(left) Ratio of dissipation energy to total turbulent energy over mixing layer height for different magnetic field strengths; (right) Variation of turbulent magnetic energy (TME) and turbulent kinetic energy (TKE) ratio with mixing layer height for different magnetic field strengths.}
    \label{C0C1}
\end{figure}

Beginning with the first assumption, we calculate the total energy dissipated up until time $t$. With the introduction of the magnetic field, the total energy dissipated $(D_E)$ is the sum of energy dissipated due to TKE and TME. Hence, the energy dissipation is calculated following equation,
\begin{equation}
    D_E = \underbrace{\int_0^t \int_V \left[ \nu \rho (\partial_j u_i)^2 + (\nu + D) (\partial_j \rho) \left(\partial_j \frac{u_i u_i}{2} \right) \right] \mathrm{d}V  \mathrm{d}t}_\text{TKE dissipation $(D_{TKE})$} + \underbrace{\int_0^t \int_V \eta (\partial_j b_i)^2 \mathrm{d}V \mathrm{d}t}_\text{TME dissipation $(D_{TME})$}.
    \label{dissipation_formula}
\end{equation}
Figure \ref{C0C1}(left) shows the ratio of energy dissipation to the total turbulent energy over mixing layer height. This ratio is referred to as $C_{diss}$. For all magnetic field strengths, $C_{diss}$ is approximately constant with mixing layer height for $h \gtrapprox 1.2$. The standard deviation for all cases is less than $2\%$. We note that $C_{diss}$ (and other coefficients) vary with magnetic field strengths. We will discuss the effect of the magnetic field on the coefficients in $\S$\ref{Mag_field_effect}.

The second assumption is the scaling of TKE and TME in the self-similar region. As before, we plot the ratio of volume-averaged TME to the volume-averaged TKE, referred to as $C_{ep}$,  over mixing layer height. Figure \ref{C0C1}(right) shows that $C_{ep}$ is approximately constant for all magnetic field strengths for $h \gtrapprox 1.2$. This is evidence that the two quantities are proportional to each other. The standard deviation for all cases is less than $1\%$.

\begin{figure}
    \centering
    \begin{subfigure}[b]{0.48\textwidth}
         \centering
         \includegraphics[width=\textwidth]{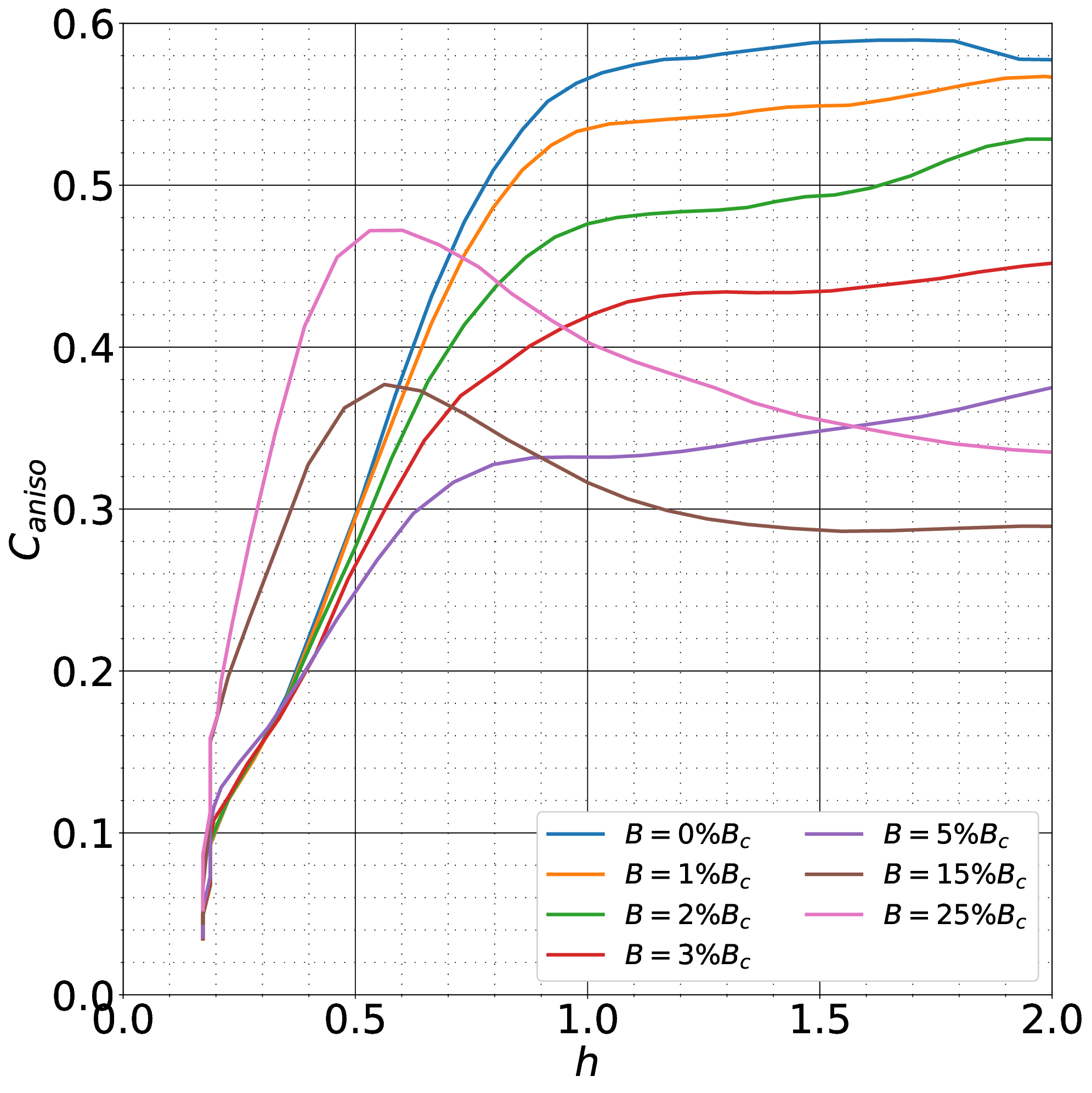}
     \end{subfigure}
     \hfill
     \begin{subfigure}[b]{0.48\textwidth}
         \centering
         \includegraphics[width=\textwidth]{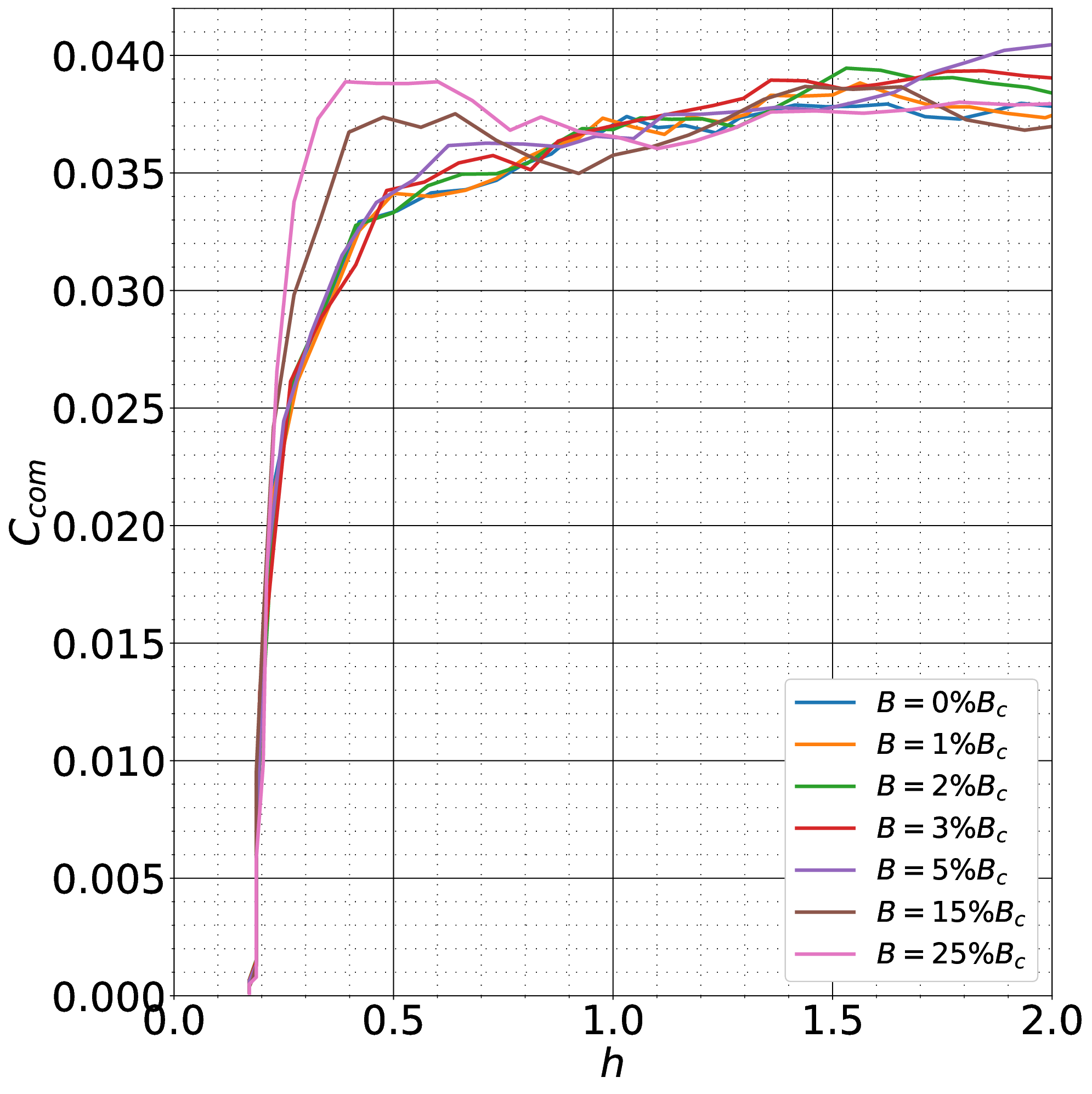}
     \end{subfigure}
    \caption{(left) Ratio of turbulent kinetic energy due to $x-$ and $y-$components of velocity to the $z-$component of velocity with mixing layer height} for different magnetic field strengths; (right) Variation of center of mass of the mixing layer ($C_{com}$, cf. equation \ref{C5B}) with mixing layer height for different magnetic field strengths.
    \label{C2C5}
\end{figure}

The third assumption is the scaling of the homogeneous and non-homogeneous kinetic energies, referred to as $C_{aniso}$. We plot the ratio of volume-averaged TKE along the homogeneous and the non-homogeneous direction against mixing layer height. From figure \ref{C2C5}(left), the ratio is approximately constant with a standard deviation less than $3\%$ for all cases. The confirmation of the above three assumptions indicates that the system achieves self-similarity beyond $h \gtrapprox 1.2$. 

Next, the released GPE $(\int_V \delta \rho g x_3 \mathrm{d}V)$ was written in terms of the mixing layer height, which gives $C_{com}$. Figure \ref{C2C5}(right) shows the variation of the quantity with mixing layer height. The $C_{com}$ is approximately constant for $h \gtrapprox 1.2$ and for all magnetic field strengths. The standard deviation of $C_{com}$ for all magnetic field strengths is less than $0.1\%$. 

\begin{figure}
    \centering
    \begin{subfigure}[b]{0.475\textwidth}
         \centering
         \includegraphics[width=\textwidth]{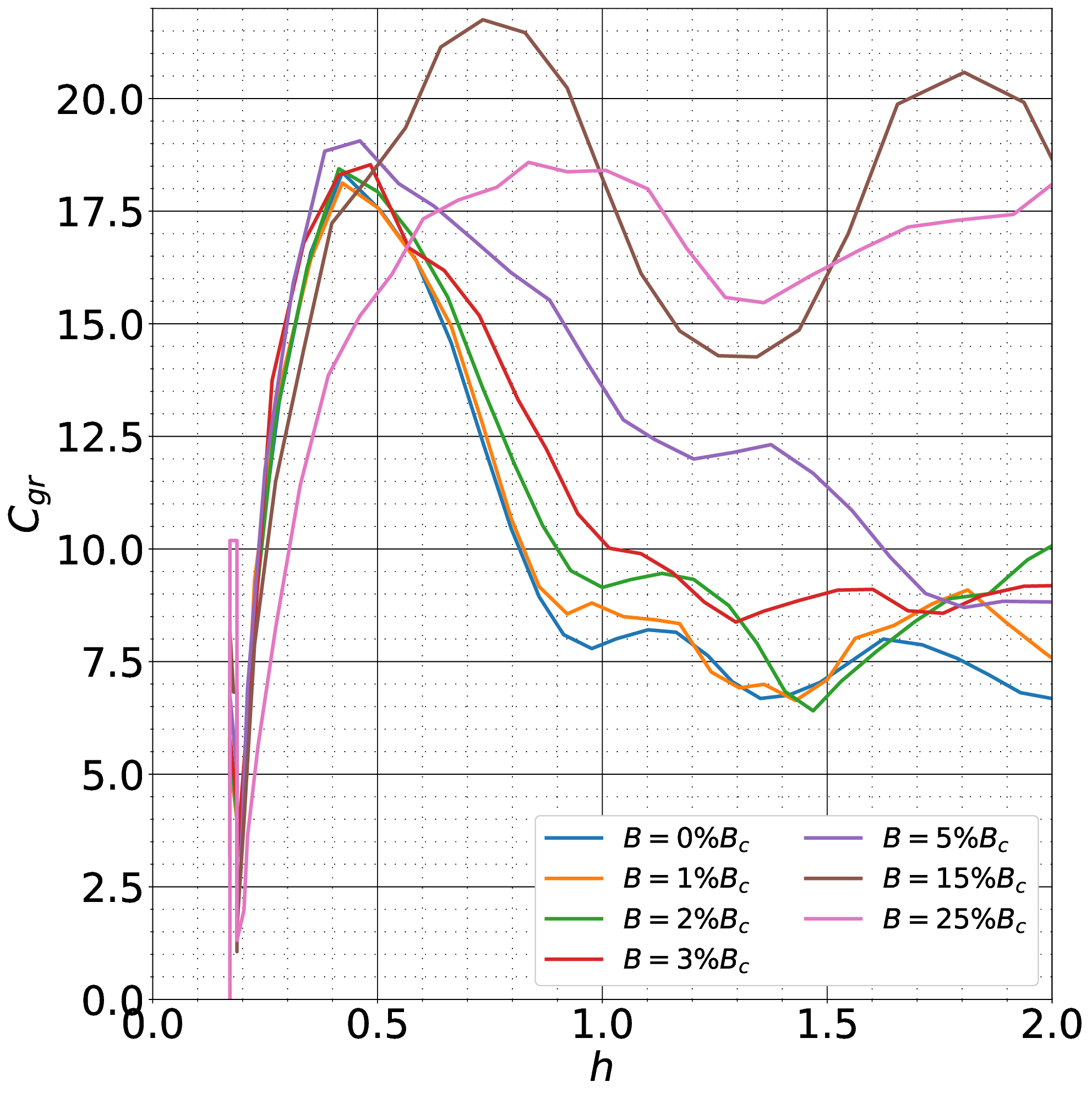}
     \end{subfigure}
     \hfill
     \begin{subfigure}[b]{0.50\textwidth}
         \centering
         \includegraphics[width=\textwidth]{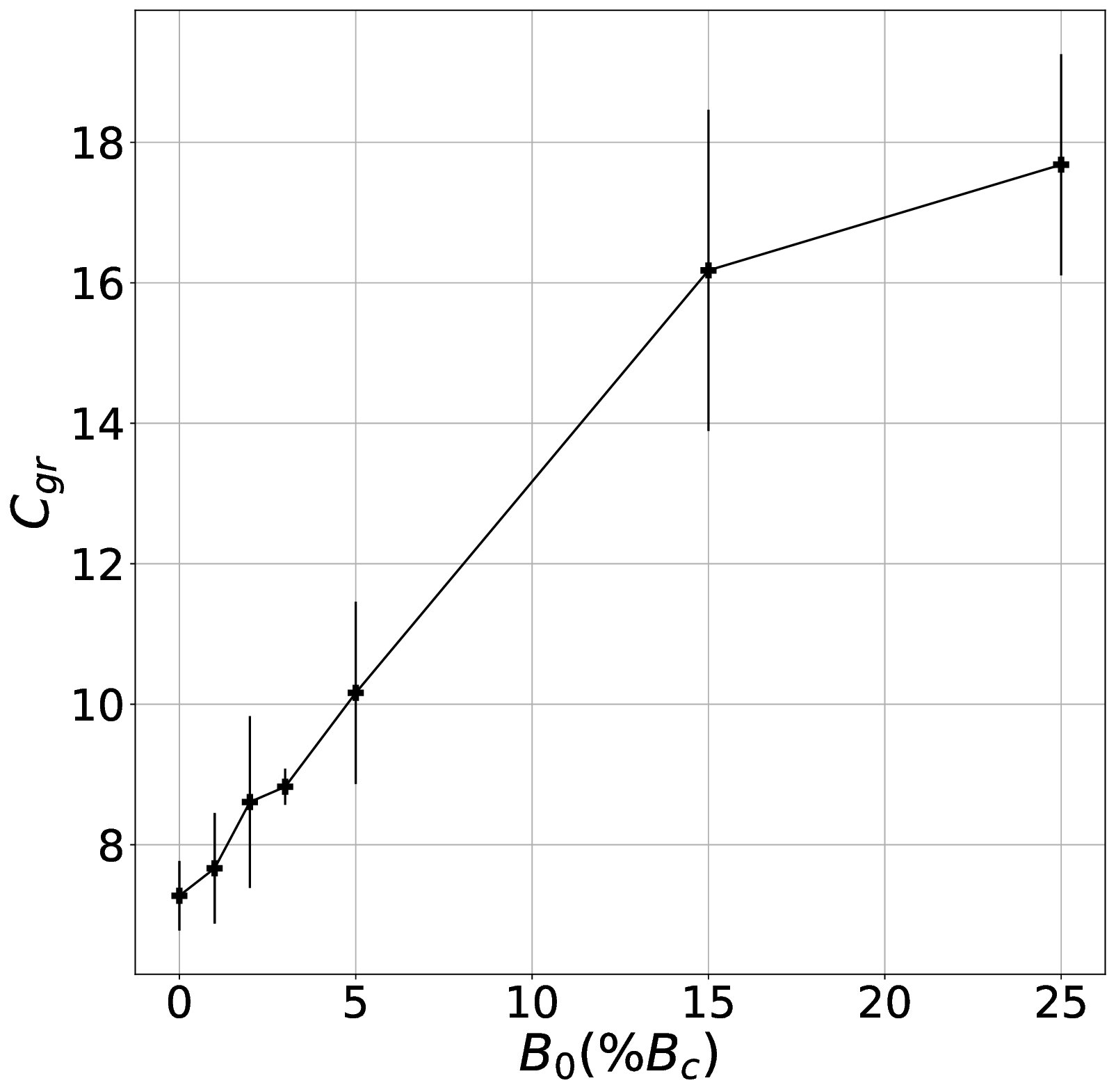}
     \end{subfigure}
    \caption{(left) Ratio of growth rate of mixing layer to the $z-$component of velocity ($C_{gr}$, cf. equation \ref{C4B}) over mixing layer height for different magnetic field strengths; (right) The variation of ($C_{gr}$) with magnetic field strength.}
    \label{C4}
\end{figure}

Lastly, towards deriving an equation of the mixing layer height, we wrote the vertical kinetic energy in terms of the mixing layer height, which gives $C_{gr}$. Physically, this quantity tells the correlation between the vertical kinetic energy and the growth rate of the mixing layer height. In figure \ref{C4}(left), we plot $C_{gr}$ with mixing layer height. The time gradient results in noisy data, and hence the data is smoothed using a rolling mean with a window of $0.4$ time units. The raw data is not shown here for legibility. The statistical mean and standard deviation of the $C_{gr}$ (in the regime $h \gtrapprox 1.2$) are shown in Figure \ref{C4} (right). 

\begin{figure}
    \centering
    \begin{subfigure}[b]{0.49\textwidth}
         \centering
         \includegraphics[width=\textwidth]{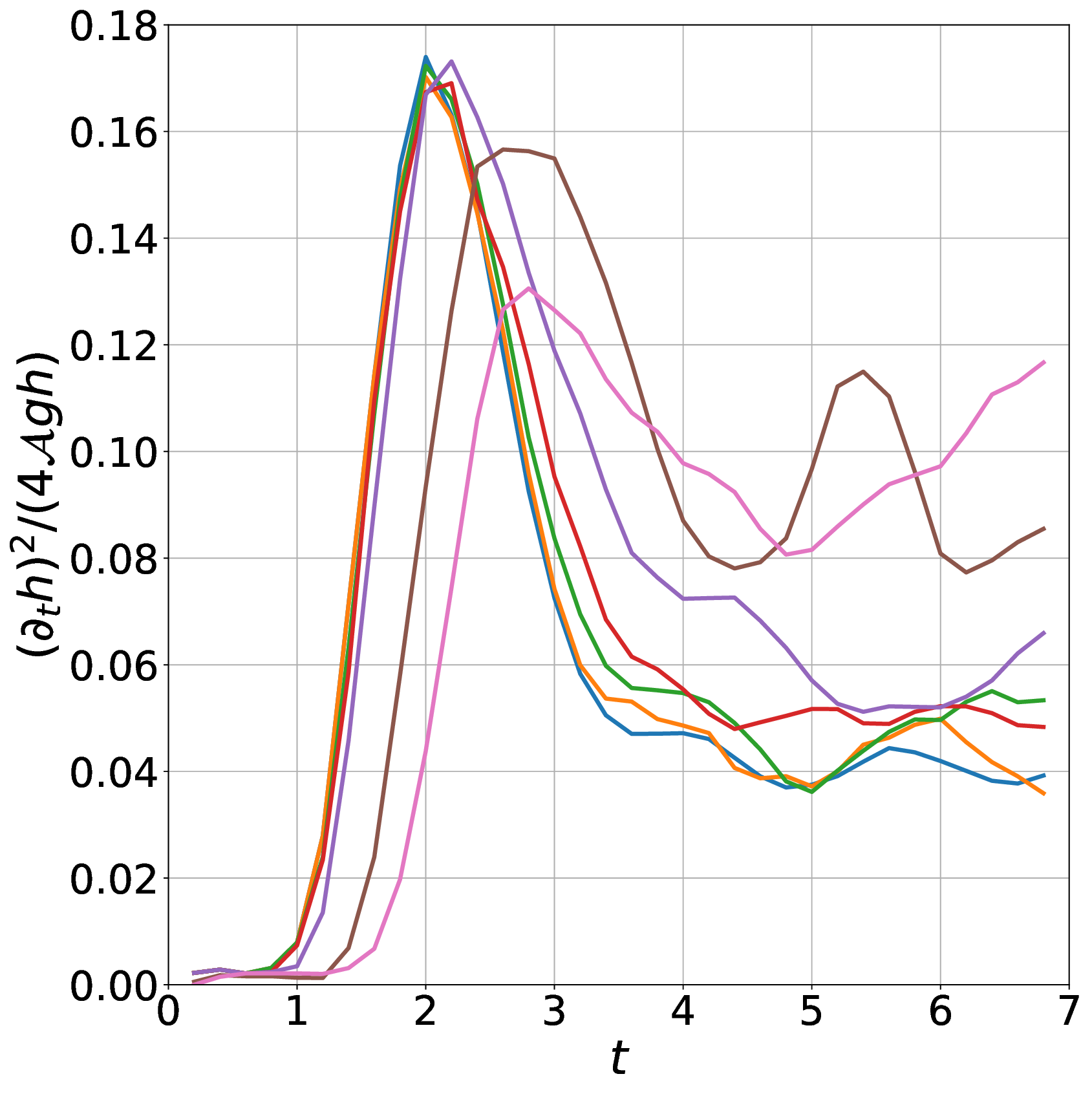}
     \end{subfigure}
     \hfill
     \begin{subfigure}[b]{0.49\textwidth}
         \centering
         \includegraphics[width=\textwidth]{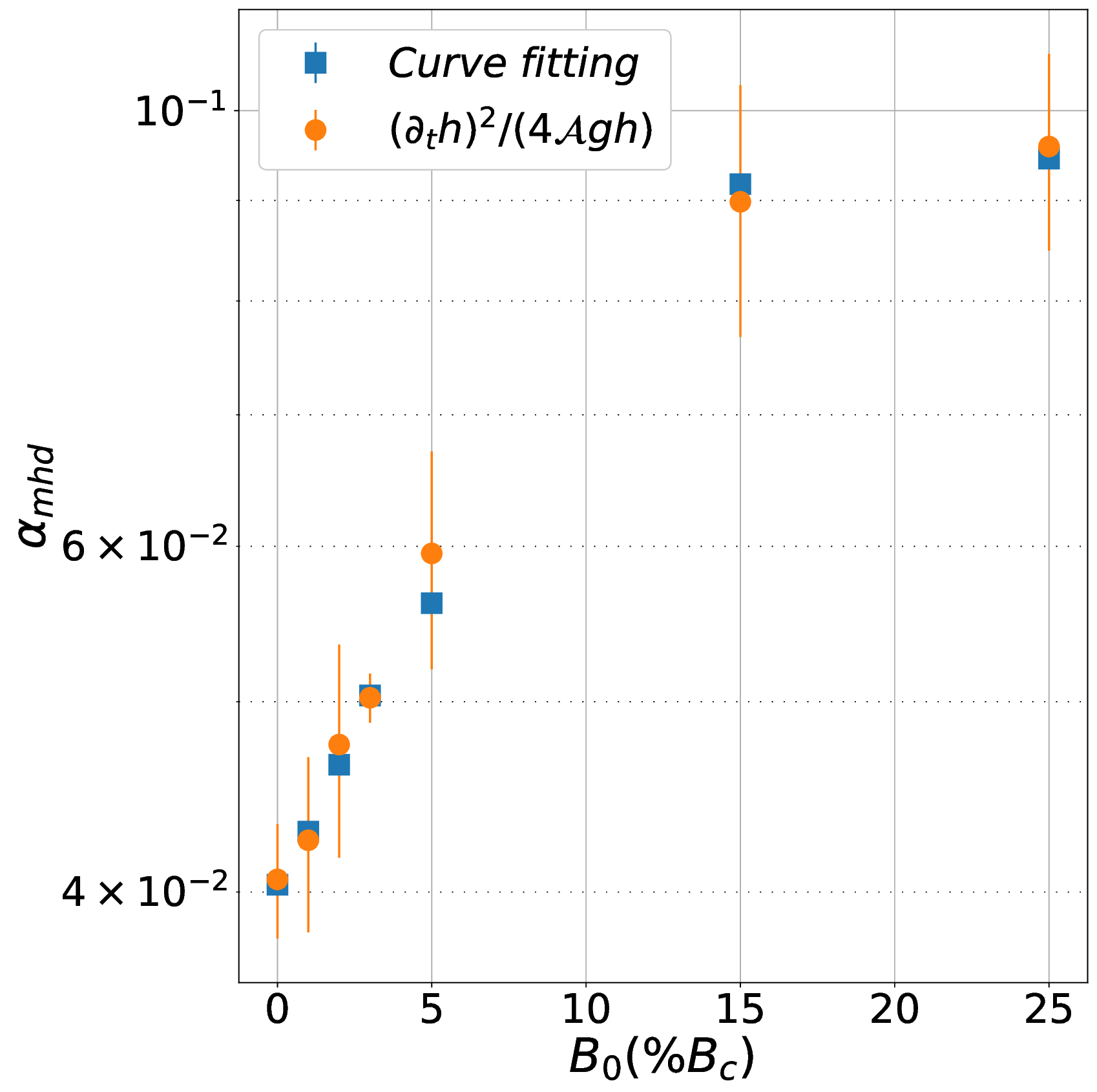}
     \end{subfigure}
    \caption{(left) Temporal variation of $\alpha_{mhd} {=} (\partial_t h)^2/4 \mathcal{A} g h$ for different magnetic field strengths; (right) Comparison of $\alpha_{mhd}$ obtained from curve fitting and mean of $(\partial_t h)^2/4 \mathcal{A} g h$ between $t \in [4, 7]$.}
    \label{dth_h}
\end{figure}

As a final confirmation of self-similarity, we plot the temporal variation of $\alpha_{mhd} = (\partial_t h)^2/4 \mathcal{A} g h$. The plot of $\alpha_{mhd}$ with $t$, shown in figure \ref{dth_h}(left), confirms that the $\alpha_{mhd}$ is approximately constant between $t = [4, 7]$ and hence the system is self-similar at $t \geq 4$. The time $t = 4$ corresponds to $h \approx 1.1$ for all magnetic field strengths except $B_0 = 25\%B_c$, see figure \ref{h_t}(left). The $\alpha_{mhd}$ is smoothed using a rolling mean with a window of 0.4 time units. The percentage of standard deviation (about the mean value) is at most $15\%$. It can be seen that the temporal variation of $\alpha_{mhd}$ mimics the temporal variation of $C_{gr}$ (see figure 5.6(left)), demonstrating that the $\alpha_{mhd}$ is predominantly influenced by the $C_{gr}$. Note that the values of $\alpha_{mhd}$ obtained from taking the mean of $(\partial_t h)^2/4 \mathcal{A} g h$ in $t \in [4, 7]$ also agree with the $\alpha_{mhd}$ estimation based on curve fitting in $\S$\ref{sec:estimation}, as shown in figure \ref{dth_h}(right).

\subsubsection{Growth rate of mixing layer} \label{growth_rate}

\begin{figure}
    \centering
    \begin{subfigure}[b]{0.49\textwidth}
         \centering
         \includegraphics[width=\textwidth]{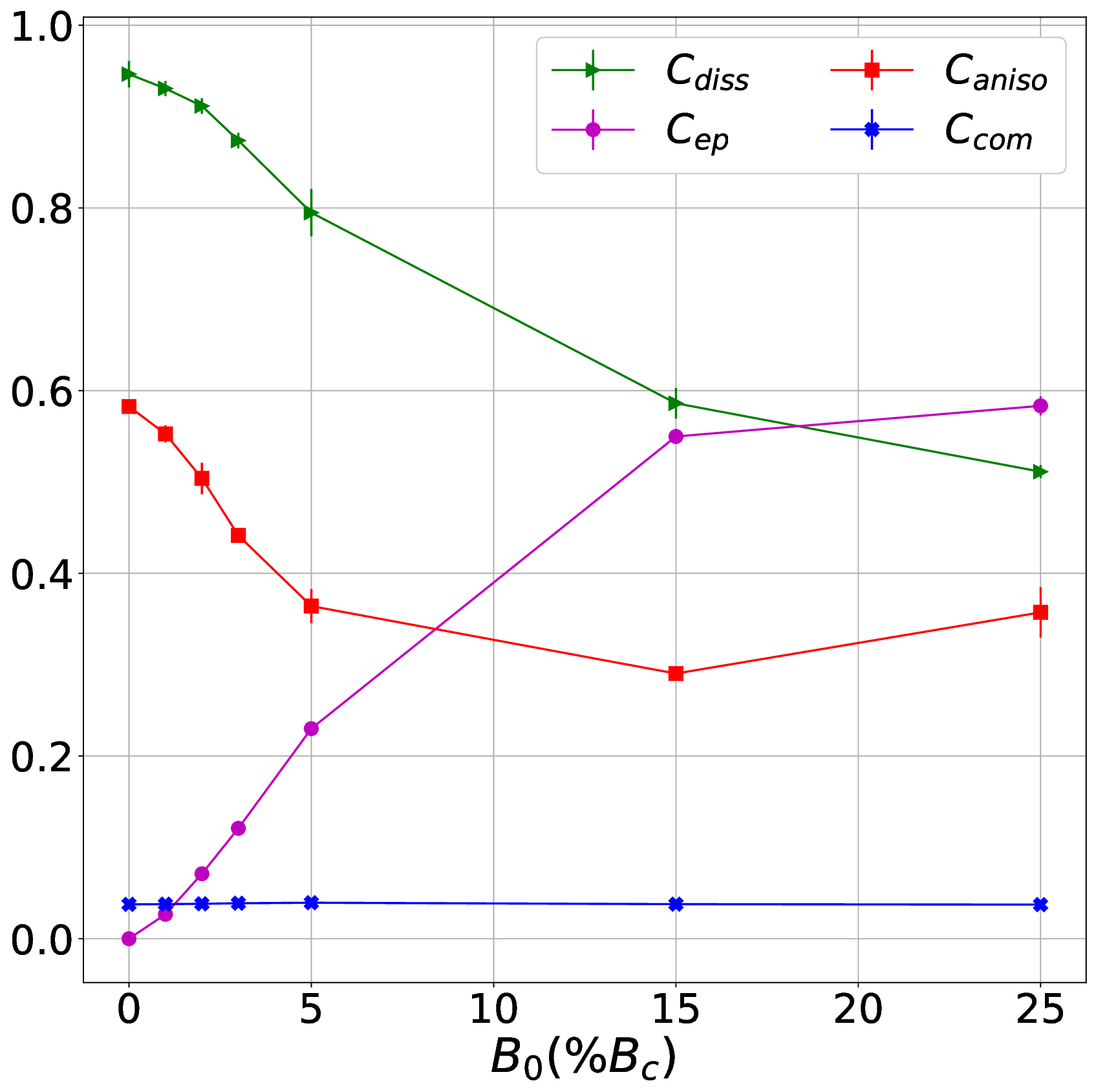}
     \end{subfigure}
     \hfill
     \begin{subfigure}[b]{0.49\textwidth}
         \centering
         \includegraphics[width=\textwidth]{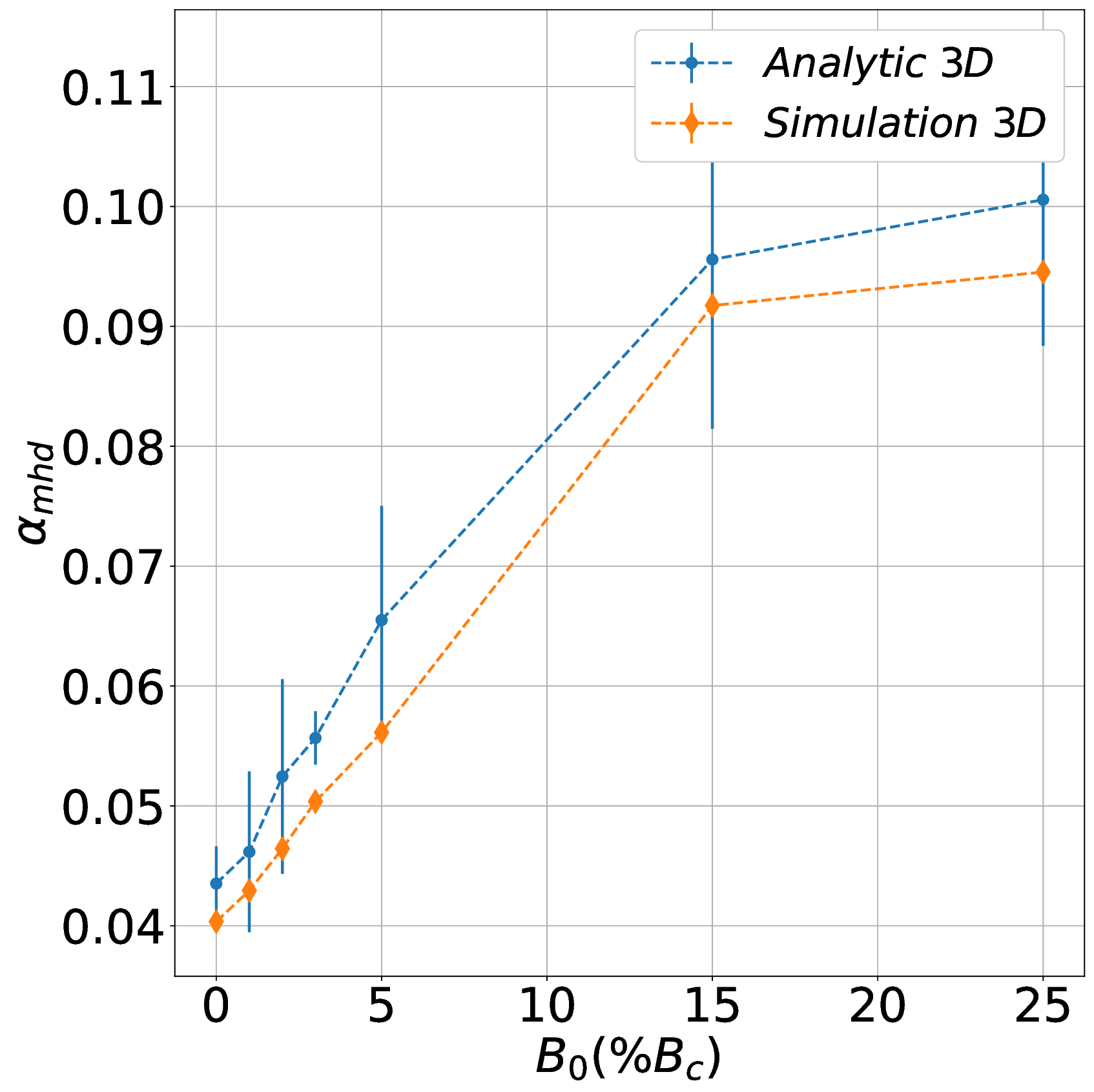}
     \end{subfigure}
    \caption{(left) Variation of different coefficients ($C_{diss}$, $C_{ep}$, $C_{aniso}$, $C_{gr}$) with magnetic field strength; (right) Comparison of $\alpha_{mhd}$ obtained from the numerical simulation against the $\alpha_{mhd}$ from analytical formula. The figure shows the $\alpha_{mhd}$ from the two formulations (analytic and numerical)}
    \label{alpha_comp}
\end{figure}

Having confirmed that the MRTI achieves self-similarity in the regime $t \gtrapprox 4$, we sample the data between $t \in [4, 7]$ and determine the scaling coefficients ($C_{diss}$, $C_{ep}$, $C_{aniso}$, $C_{gr}$, $C_{com}$) by taking their mean in the above time interval. The mean and standard deviation values of $\alpha_{mhd}$ is calculated as below (equations \ref{mean_alpha}, \ref{std_alpha}) cf. the formula \ref{intermediate}:
\begin{equation}
    \overline{\alpha}_{mhd} = \frac{\overline{C_{com}} \text{\hspace{5pt}} \overline{C_{gr}}}{2 (1 {+} \overline{C_{diss}}) (1 {+} \overline{C_{ep}}) (1 {+} \overline{C_{aniso}})},
    \label{mean_alpha}
\end{equation}
\begin{equation}
    \sigma(\alpha_{mhd}) = \overline{\alpha}_{mhd} \left( \frac{\sigma(C_{diss})}{\overline{C_{diss}}} {+} \frac{\sigma(C_{ep})}{\overline{C_{ep}}} {+} \frac{\sigma(C_{aniso})}{\overline{C_{aniso}}} {+} \frac{\sigma(C_{gr})}{\overline{C_{gr}}} {+} \frac{\sigma(C_{com})}{\overline{C_{com}}} \right).
    \label{std_alpha}
\end{equation}
Here $\sigma(\star)$, $\overline{\star}$ represents the standard deviation and mean of $\star$. The variation of each coefficient and $\alpha_{mhd}$ with the imposed magnetic field is shown in Figure \ref{alpha_comp} (for $C_{gr}$, see Figure \ref{C4}(right)). The markers show the mean values, and the error bar shows the standard deviation. The uncertainty in $\alpha_{mhd}$ is predominantly due to the growth rate coefficient, $C_{gr}$.

For the sake of comparison, in figure \ref{alpha_comp} (right), we included the $\alpha_{mhd}$ calculated from numerical simulations. We see that the trend of $\alpha_{mhd}$ from the numerical simulations and analytical calculations are consistent. The value of $\alpha_{mhd}$ from numerical simulations falls within the error bars of $\alpha_{mhd}$ obtained analytical formula. However, the $\alpha_{mhd}$ obtained from the analytical formula seems to overestimate the non-linear growth from the numerical simulations marginally. This could be due to the choice of mixing layer height definition. In the current study, the mixing layer height is calculated based on the threshold of the mixing layer parameter $\Theta = 0.1$. Choosing a different threshold of $\Theta$ or using an integral definition for mixing layer height are known to influence the mixing layer height \citep{baltzer_livescu_2020}. Since $C_{gr}$ and $C_{com}$ depend on mixing layer height (particularly $C_{gr}$ is sensitive due to temporal gradient, $\partial_t h$), the value of $\alpha_{mhd}$ also changes. 

\subsection{Effect of magnetic field} \label{Mag_field_effect}

\subsubsection{Variation of the coefficients with magnetic field strength}

\begin{figure}
    \centering
    \begin{subfigure}[b]{0.46\textwidth}
         \centering
         \includegraphics[width=\textwidth]{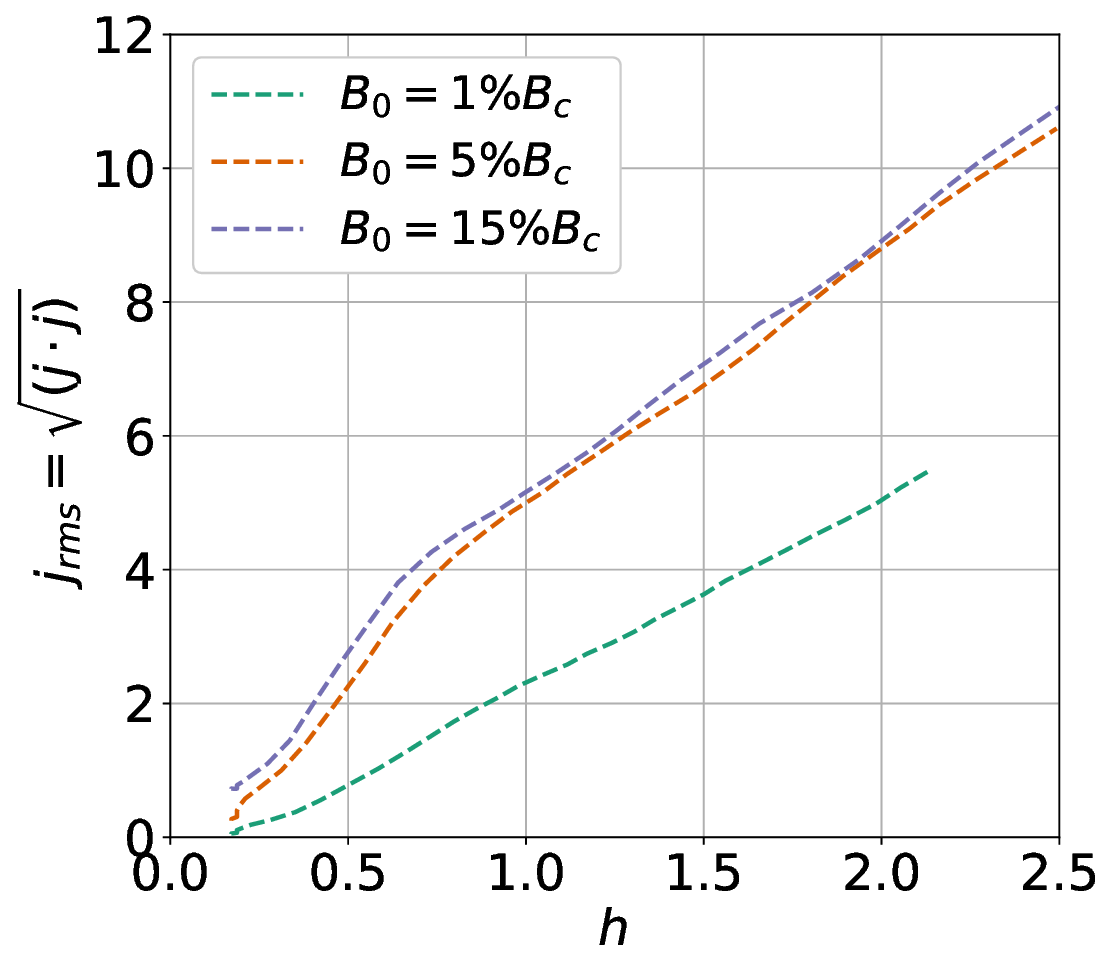}
     \end{subfigure}
     \hfill
     \begin{subfigure}[b]{0.46\textwidth}
         \centering
         \includegraphics[width=\textwidth]{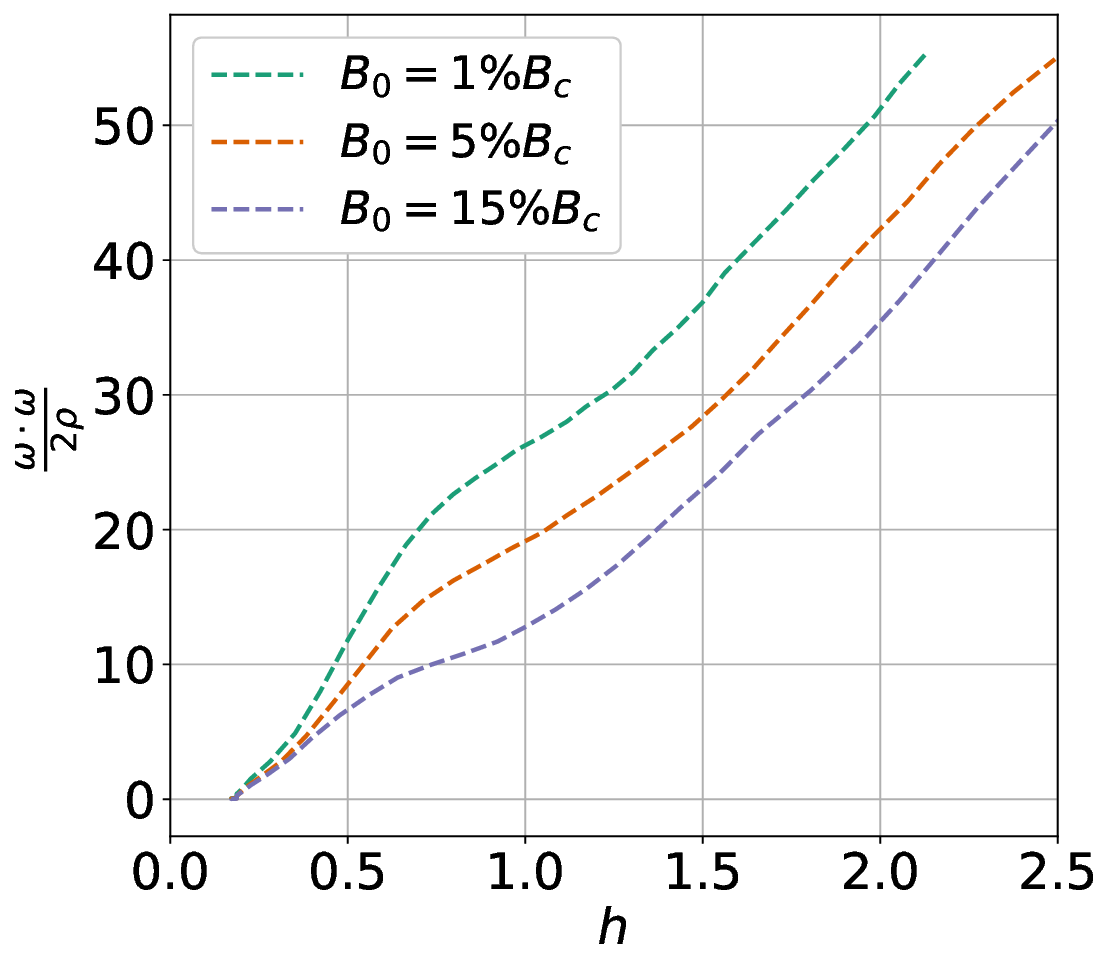}
     \end{subfigure}
    \caption{Temporal variation of (left) root mean square current, (right) root mean square vorticity for different magnetic field strengths.}
    \label{currentvorticity}
\end{figure}

\begin{figure}
    \centering
    \begin{subfigure}[b]{0.46\textwidth}
         \centering
         \includegraphics[width=\textwidth]{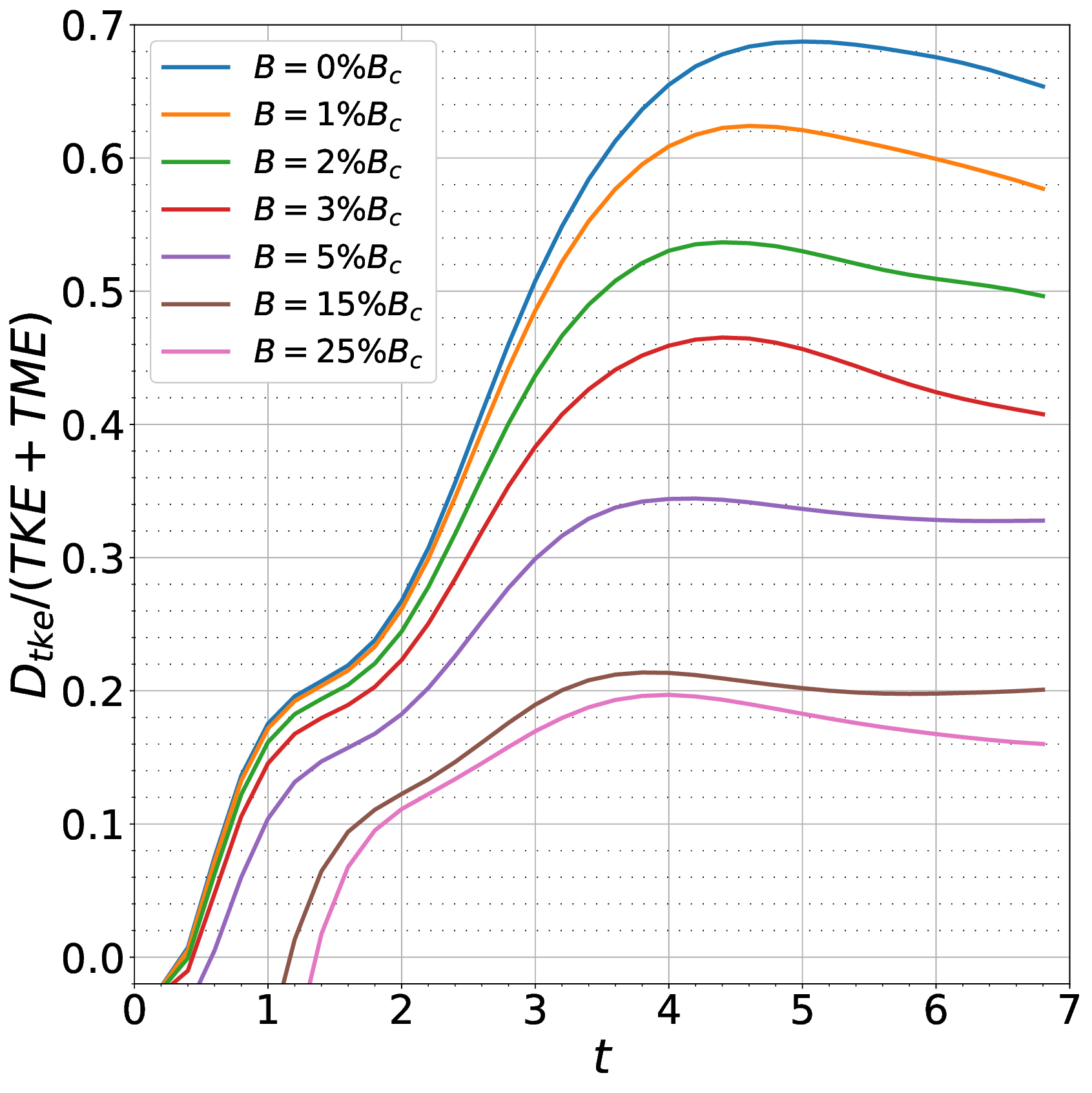}
     \end{subfigure}
     \hfill
     \begin{subfigure}[b]{0.46\textwidth}
         \centering
         \includegraphics[width=\textwidth]{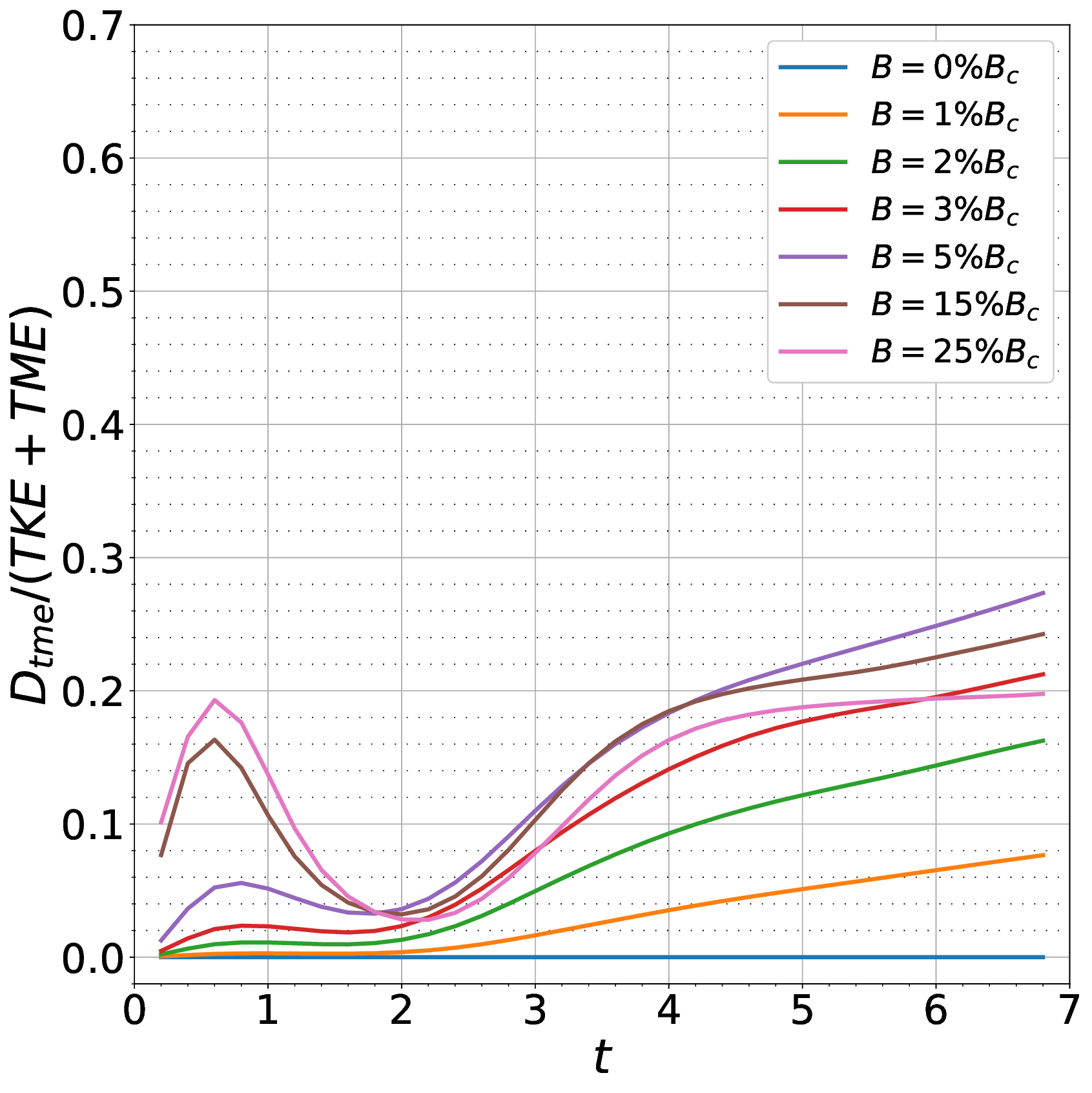}
     \end{subfigure}
    \caption{Temporal variation of (left) turbulent kinetic energy dissipation, (right) turbulent magnetic energy dissipation for different magnetic field strengths. The energy dissipation is normalized with the total turbulent energy.}
    \label{Dtke_Dtme}
\end{figure}

The figures \ref{alpha_comp}(left) and \ref{C4}(right) show that the quantities $C_{diss}$, $C_{ep}$, $C_{aniso}$, $C_{gr}$ vary significantly with the magnetic fields. But the reason for this variation is not discussed. This is the focus of the present section. Starting with $C_{diss}$ $\left( \textcolor{ForestGreen}{-{\blacktriangleright}-} \right)$, the dissipation is highest for the HD case ($\approx 67.5\%$ of the total turbulent energy or $\approx 40\%$ of the released GPE). With the introduction of the magnetic field, the magnetic field suppresses the small wavelength perturbations (cf. $\S$\ref{intro}) and the vorticity ($\omega$), thus leading to lower TKE dissipation ($D_{tke} \propto \nu \omega^2$ \citep{Bandyopadhyay2018}). But the introduction of magnetic fields also introduces the current sheet, which dissipates, resulting in TME dissipation ($D_{tme} \propto \eta j^2$ \citep{Bandyopadhyay2018}, $j$ is the strength of the current sheet). As the strength of the imposed magnetic field increases, the strength of the current sheets increases, and enstrophy in the system decreases, see figure \ref{currentvorticity}. Thus, we expect an increasing TME dissipation and decreasing TKE dissipation with increasing magnetic field strength. Yet, the total energy dissipation ($C_{diss}$) seems to decrease with increasing magnetic field (see figure \ref{alpha_comp}(left)). To understand this, we plot the energy dissipation due to TKE and TME in Figure \ref{Dtke_Dtme}. The energy dissipation is normalized with the total turbulent energy. With the increase in magnetic field strength, the energy dissipation due to TKE reduces significantly, but the increase in TME dissipation is marginal. This leads to a net decrease in energy dissipation with increasing magnetic field strength. Similar observations were made in previous studies \citep{Bandyopadhyay2018}. Further, figure \ref{Dtke_Dtme}(right) shows that the TME dissipation increases up to $5\% B_c$ and then decreases.

\begin{figure}
    \centering
    \includegraphics[width= 0.9\linewidth]{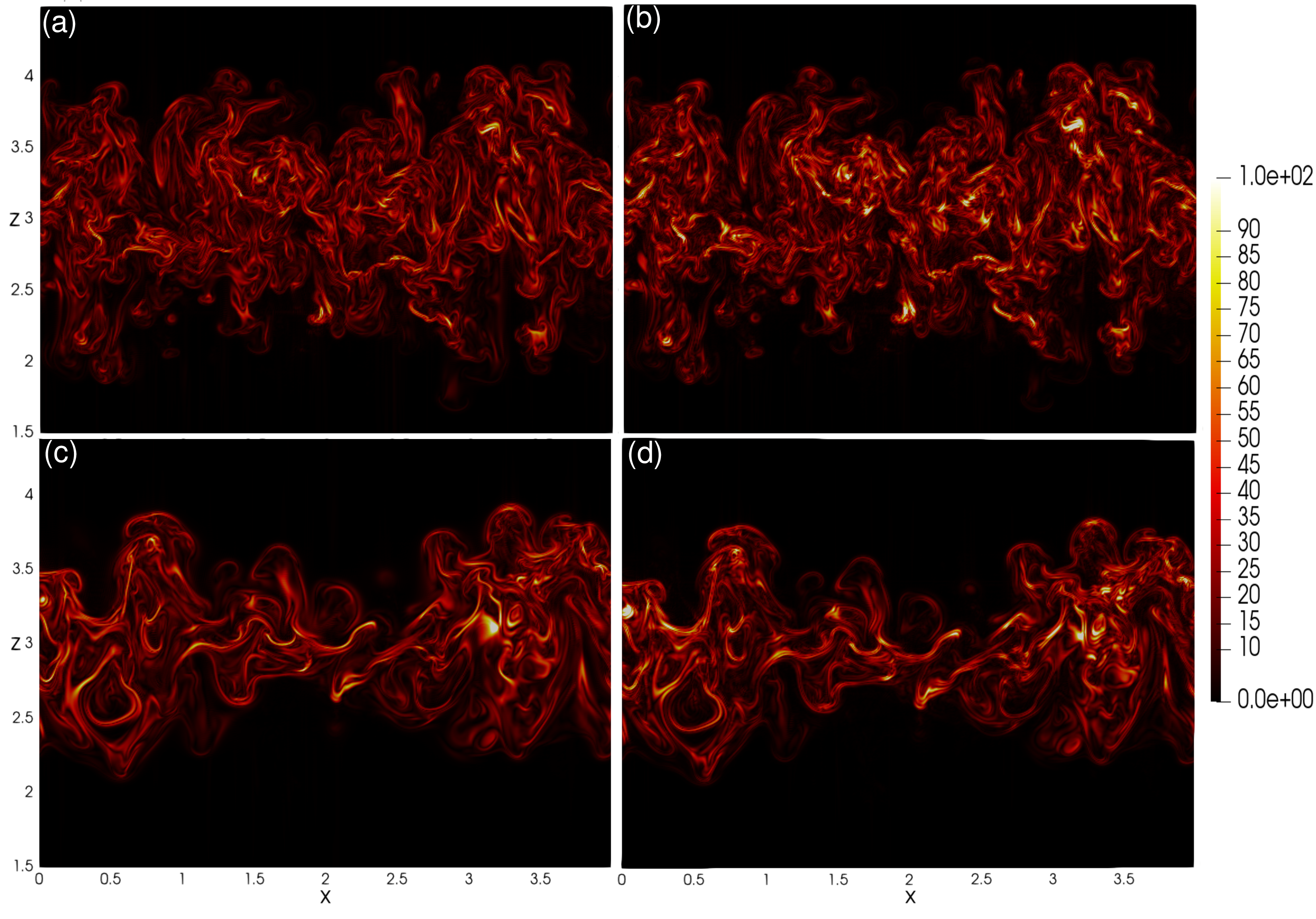}% vorticity_current.eps}
    \caption{Instantaneous 2D contours of 3D MRTI. The left panels (a, c) shows current, and the right panels (b, d) shows vorticity. The top and bottom panels correspond to $B_0 = 5\%B_c, 25\%B_c$ cases, respectively.}
    \label{vorticity_current}
\end{figure}

\begin{figure}
    \centering
    \includegraphics[width=0.45\textwidth]{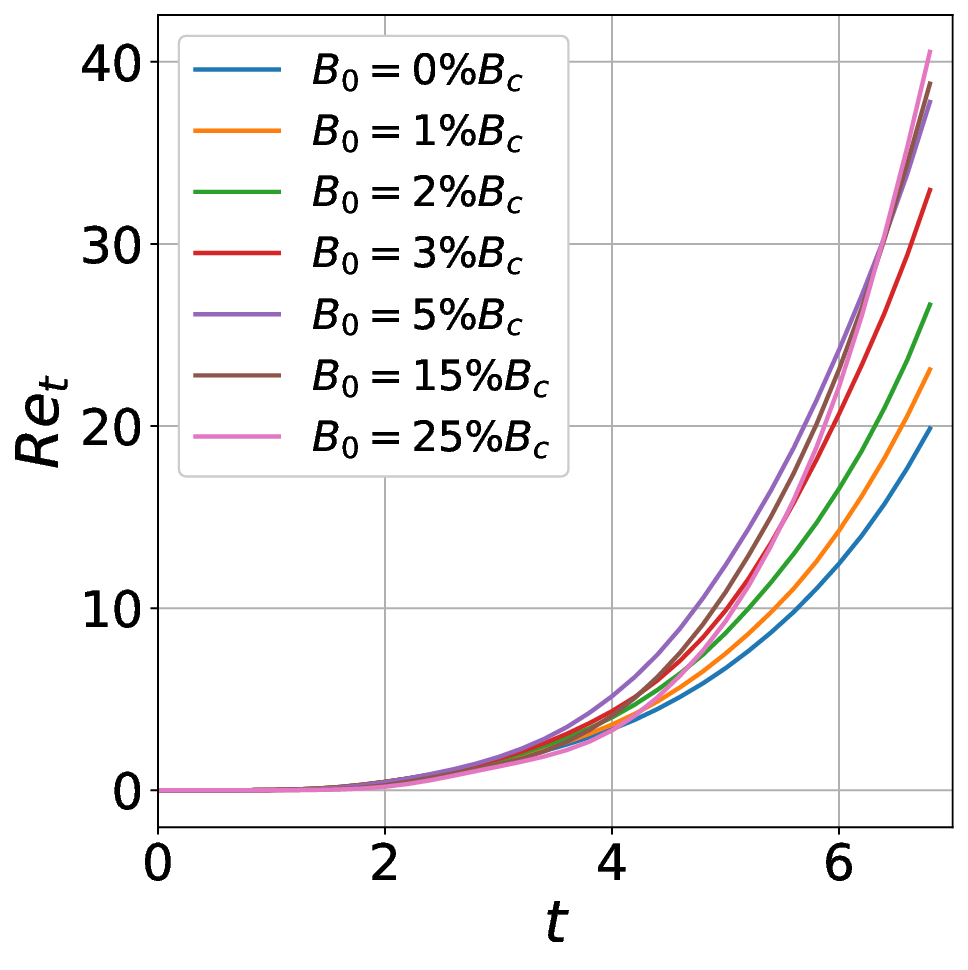}
    \caption{Temporal variation of turbulent Reynolds number for different magnetic field strengths}
    \label{Re}
\end{figure}

Why does the energy dissipation reduce significantly with the introduction of the magnetic field? The two important reasons for this. First, the magnetic field (particularly strong $B_0$) inhibits the development of small-scale structures (like current sheets and vortices) in directions parallel to the field. This is demonstrated in figure \ref{vorticity_current} where the current (left panel) and vorticity (right panel) have more small-scale structures in the weak field case $B_0 = 5\% B_c$ (top panel), compared to the strong magnetic field case $B_0 = 25\%B_c$ (bottom panel). Since dissipation occurs at small scales where gradients are steep. The suppression of small-scale structures reduces the overall dissipation. Second, as the mean magnetic field strength increases, the structures become larger (as shown in figure \ref{isocontours}), increasing the turbulent Reynolds number, see figure \ref{Re}, and decreasing the dissipation. The turbulent Reynolds number is defined as $Re_t = \frac{\overline{\rho} K^2}{\nu \varepsilon}$, where $\overline{\rho} = \frac{\rho_h + \rho_l}{2}$, $K = \frac{1}{V_m} \int_{V_m} \sum_{i = 1}^3 u_i^2 \mathrm{d}V_m$, and $\varepsilon = \frac{1}{V_m} \int_{V_m} \nu \rho (\partial_j u_i)^2 \mathrm{d}V_m$.

Contrary to $C_{diss}$, the $C_{ep}$ $\left( \textcolor{Purple}{-{\CIRCLE}-} \right)$ is a monotonically increasing parameter. As the imposed field strength increases, the amount of energy required to deform the magnetic field lines increases. Hence, a greater proportion of energy is invested in TME energy. Therefore, the magnetic energy increases with magnetic field strength. However, $C_{ep}$ seems to increase rapidly in the weak magnetic field case (up to $B_0 \approx 5\%B_c$). Beyond this, the increase is less steep. A potential reason for this is the significant deformations of magnetic field lines at a wide range of scales in the weak magnetic field case. This is evident from the figure \ref{vorticity_current}(b). In the strong magnetic field case, most of the small scales are suppressed by the magnetic field (see $\S$\ref{intro}), restricting the deformation of field lines to large scales. 

$C_{aniso}$$\left( -{\blacksquare}- \right)$ was also found to decrease with increasing magnetic field strength. That is, the proportion of TKE in the homogeneous direction decreases with increasing field strength. A potential reason for this is the reduction of vortical structures with increasing field strength (see figures \ref{vorticity_current} (b, d)). Typically, the released GPE is injected into the system through the non-homogeneous component of TKE. The vorticity acts to redistribute the energy from non-homogeneous to homogeneous directions (for example, through vortex stretching) \citep{Saffman_1993}. At large magnetic fields, due to the reduction of the vortical structures (see figure \ref{currentvorticity}(right)), the energy is mostly in the non-homogeneous direction, increasing anisotropy (i.e., decreasing $C_{aniso}$). The increasing difference in TKE between the homogeneous and non-homogeneous directions with increasing magnetic field strength was previously observed in the MHD turbulence \citep{Shebalin_Matthaeus_Montgomery_1983}. Under the assumption perfect energy distribution across all directions $\left(\frac{1}{2} \rho u_1^2 = \frac{1}{2} \rho u_2^2 = \frac{1}{2} \rho u_3^2 \right)$, the $C_{aniso}$ tends to a maximum value of $2$. The current HD case shows significant energy redistribution with $C_{aniso} \approx 0.585$, i.e., $\approx 29\%$ of ideal redistribution.

With the absence of small-scale vortical structures (that redistribute energy into other components of TKE), the vertical velocity is primarily invested towards the growth of the mixing layer height in the strong magnetic field case. Hence, we expect an increasing value of the $C_{gr}$ with increasing field strength. This is exactly seen in the numerical simulations, figure \ref{C4}(right). 

The center of mass of the mixing layer does not vary significantly with the field strength ( $C_{com} \approx 0.0385$). This contradicts the results of \citet{Carlyle2017}, where they found that the center of mass of the mixing layer changes with the field strength. However, \citet{Carlyle2017} used a much larger density contrast, $\rho_h = 10 \rho_l$. 

\subsubsection{Scaling laws of coefficients} \label{scalings}

The variation of the coefficients ($C_{diss}$, $C_{ep}$, $C_{aniso}$, $C_{gr}$, $C_{com}$) with magnetic field strength discussed previously brings the question, do these coefficients follow a scaling law with imposed magnetic field strength? Such a scaling relation, if obtained, could be useful towards interpolating the coefficients (and thus the non-linear growth constant) at any intermediate magnetic field strength, without simulating MRTI. Towards this, we first develop the scaling laws building from the previous studies, and verify if the current results match the proposed scaling laws.

Towards this direction, we would first look at the energy dissipation. The energy dissipation is the combination of energy dissipation due to turbulent kinetic and magnetic energies, which can be assumed as $\eta j^2$ and $\nu \rho \omega^2$ \citep{Bandyopadhyay2018}, where $j$ is the current, $\omega$ is the vorticity. The quantities $\eta j^2$ and $\nu \rho \omega^2$ can be assumed to scale as $\eta b_{rms}^2/l_j^2$ and  $\nu \rho u_{rms}^2/l_{\omega}^2$, respectively. Here $l_j$ and $l_{\omega}$ are the length scales of characteristic current sheets and characteristic vorticity, $b_{rms}$ and $u_{rms}$ correspond to RMS values of turbulent magnetic field and turbulent velocity. Hence, the total energy dissipation is proportional to the sum of $\eta b_{rms}^2/l_j^2$ and  $\nu \rho u_{rms}^2/l_{\omega}^2$, i.e.,
\begin{equation}
    D_E \propto \left( \eta \int_0^t \int_V \frac{b_{rms}^2}{l_j^2} \mathrm{d}V \mathrm{d}t + \nu \int_0^t \int_V \frac{\rho u_{rms}^2}{l_{\omega}^2} \mathrm{d}V \mathrm{d}t \right).
\end{equation}
The above equation can be written in terms of energy partition ($C_{ep}$, cf. $\S$\ref{h-derivation}) and magnetic Prandtl number $(Pr_m)$ as,
\begin{equation}
    \implies D_E \propto \left( \frac{l_{\omega}^2}{l_j^2} \frac{1}{Pr_m} C_{ep} + 1 \right) \int_0^t \int_V \nu \frac{\rho u_{rms}^2}{l_{\omega}^2} \mathrm{d}V \mathrm{d}t, \text{ where  } C_{ep} = \frac{\int_V b_{rms}^2 \mathrm{d}V}{\int_V \rho u_{rms}^2 \mathrm{d}V}, Pr_m = \frac{\nu}{\eta}.
    \label{C0B_scaling1}
\end{equation}

From the above equation, to obtain the scaling relation for the energy dissipation, we need to obtain the scaling relation for the energy partition between the turbulent magnetic and kinetic energies first. The energy gained by the magnetic field depends on the range of the scales suppressed by the magnetic field. Thus, the energy partition between the TME and TKE is adequately obtained from the linear analysis. In the linear regime, \citet{Hillier2016} showed that the energy partition between the turbulent magnetic energy and turbulent kinetic energy varies quadratically with imposed magnetic field strength. Further, a recent study by \cite{Lazarian_2025} showed that, in an incompressible turbulent system, the ratio of TME to TKE varies quadratically with imposed magnetic field strength, i.e.,
\begin{equation}
    C_{ep} \equiv \frac{\int_V b_{rms}^2 \mathrm{d}V}{ \int_V \rho u_{rms}^2 \mathrm{d}V} \propto B_0^2.
    \label{C1B_scaling1}
\end{equation}
However, note that the \cite{Lazarian_2025} considers a fully homogeneous turbulence system, different from the turbulence in RTI where there is unstable density stratification.

Substituting equation \ref{C1B_scaling1} into equation \ref{C0B_scaling1}, we obtain the scaling law for energy dissipation as below:
\begin{equation}
    D_E \propto \left( \frac{l_{\omega}^2}{l_j^2} \frac{1}{Pr_m} a B_0^2 + 1 \right) \int_0^t \int_V \nu \frac{\rho u_{rms}^2}{l_j^2} \mathrm{d}V \mathrm{d}t.
\end{equation}
where $a$ is a constant of proportionality. Thus, we expect the energy dissipation to scale quadratically with $B_0$. In the hydrodynamic limit ($B_0 = 0$), energy dissipation reduces to dissipation due to TKE as expected. Dividing both sides of the above with total turbulent energy (TKE $+$ TME), we can show that
\begin{equation}
    C_{diss} \propto \left( \frac{l_{\omega}^2}{l_j^2} \frac{1}{Pr_m} a B_0^2 + 1 \right) C_{diss_{B=0}}.
    \label{C0B_scaling}
\end{equation}

\begin{figure}
     \centering
     \begin{subfigure}[b]{0.49\textwidth}
         \centering
         \includegraphics[width=\textwidth]{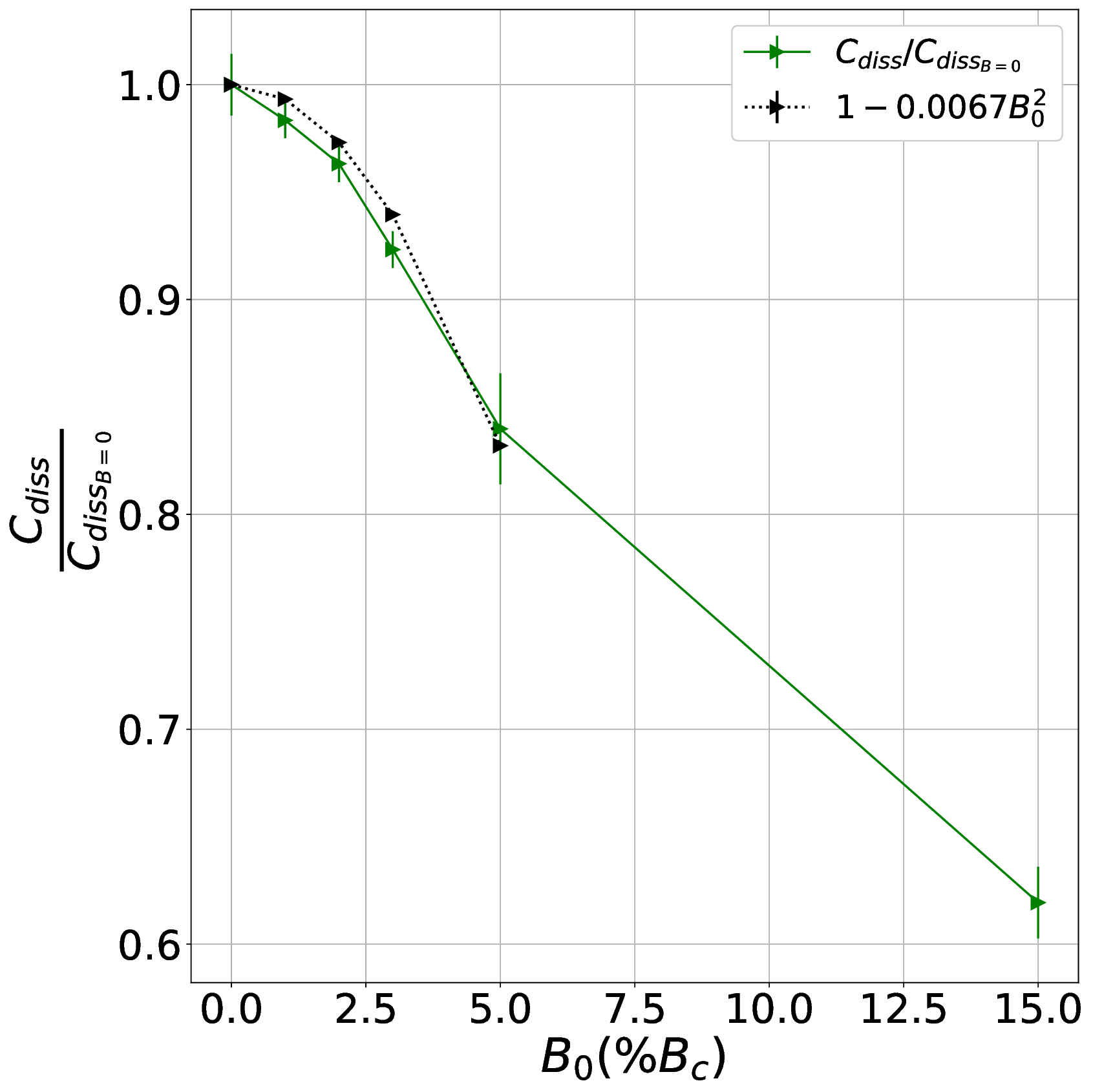}
     \end{subfigure}
     \hfill
     \begin{subfigure}[b]{0.49\textwidth}
         \centering
         \includegraphics[width=\textwidth]{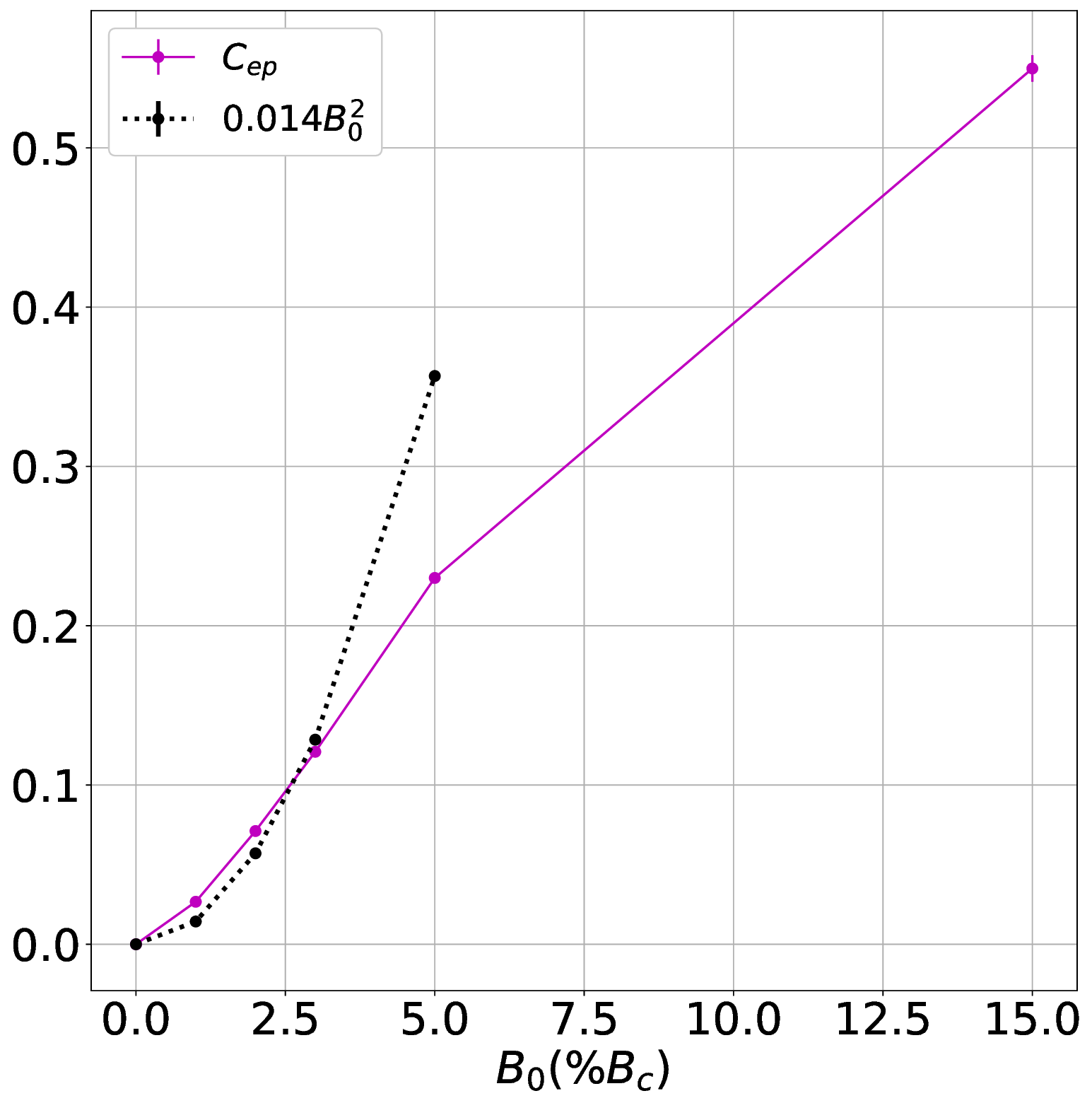}
     \end{subfigure}
     \caption{Variation of $C_{diss}$ (left) and $C_{ep}$ (right) with increasing magnetic field strength. Figure also shows the comparison of the variation of $C_{diss}$ and $C_{ep}$ with $B_0$ based on the scaling laws reported in equations \ref{C0B_scaling} and \ref{C1B_scaling1}.}
     \label{scalings1}
\end{figure}
The above theories and scaling laws apply only to the turbulent mixing layers ($B_0 < 5\% B_c$, see figure \ref{scalings1}). In the case of stronger magnetic fields, where the mixing layer is characterized by large-scale plumes alone, the above arguments do not apply and we might have to seek help from other models like the buoyancy-drag model \citep{SCHILLING2020, YOUNGS2020} to explain the variation of these quantities with magnetic field. 

\section{Discussion}

Towards validating the appropriateness of the HDRTI scaling to the MRTI, we performed an analytical self-similar analysis using the hydrodynamic (HD) scaling --- $h \propto \mathcal{A}gt^2, u \propto \mathcal{A}gt, b \propto \mathcal{A} g t \sqrt{\rho},$ and $p \propto \rho (\mathcal{A} g t)^2$. Through proof by contradiction we demonstrated that MRTI does not naturally possess the HD self-similarity, but it can converge towards the HD RTI self-similar evolution and follows the HD scaling, but only when the influence of imposed magnetic field (which deviates the system from the HD scaling) is dominated by other non-linear terms (cf. $\S$\ref{self-similarity}). It was confirmed that the influence of the imposed magnetic field term decays as $1/t$ relative to other non-linear terms (see figure \ref{B0udb_rhov}). Thus, the HD RTI self-similarity is expected to become appropriate for MRTI at late times. Further evidencing the relevance of HD scaling, in $\S$\ref{h-derivation} the mixing layer height ($h$) equation was derived analytically for the non-linear MRTI, and the equation demonstrated the quadratic growth of $h$ in the self-similar regime through energy arguments. The formula of $\alpha_{mhd}$ (equation \ref{intermediate}) highlight various physical quantities that influence the non-linear growth of the instability --- the energy dissipation ($C_{diss}$), energy partition between the turbulent magnetic and kinetic energies ($C_{ep}$), energy distribution between the homogeneous and non-homogeneous components ($C_{aniso}$), and the mixing layer growth rate per unit non-homogeneous turbulent kinetic energy ($C_{gr}$). 

\subsection{Extending the results to vertical magnetic field}
In the current study, we considered a uniform magnetic field parallel to the plane of the interface, $\mathbf{B_0} = (B_0, 0, 0)$. The theory of self-similarity in $\S$\ref{self-similarity} can be extended to the case of a vertical magnetic field, $\mathbf{B_0} = (0, 0, B_0)$. In this case, the equation of TKE in terms of the non-dimensional variables and the self-similar variable is
\begin{equation}
    \begin{split}
     \xi \langle \Tilde{\rho} \Tilde{u}_i \Tilde{u}_i \rangle {-} \xi^2 \partial_\xi \left\langle  \Tilde{\rho} \frac{\Tilde{u}_i \Tilde{u}_i}{2} \right\rangle {+} \frac{1}{2} \partial_\xi \left\langle \Tilde{u}_3 \Tilde{\rho} \frac{\Tilde{u}_i \Tilde{u}_i}{2} \right\rangle {-} \frac{1}{2} \partial_\xi \langle \Tilde{u}_3 \Tilde{p} \rangle {-} \frac{1}{2 \sqrt{\rho_m} \mathcal{A} g t} B_0 \langle \Tilde{u}_i \partial_{\xi} \Tilde{b}_i \rangle {-} \frac{1}{2} \left \langle \Tilde{u}_i \partial_{\xi} \left ( \frac{\Tilde{b}_{1} \Tilde{b}_i}{C_1} {+} \frac{\Tilde{b}_{2} \Tilde{b}_i}{C_2} {+} \Tilde{b}_{3} \Tilde{b}_i \right) \right \rangle \\ {-} \xi \frac{1}{\mathcal{A}}  \langle \Tilde{\rho} \Tilde{u}_i \delta_{i3} \rangle  = 0.
    \end{split}
\end{equation}
The equation of TME in terms of the non-dimensional variables and the self-similar variable is
\begin{equation}
    \begin{split}
         \xi \langle \Tilde{b}_i \Tilde{b}_i \rangle {-} \xi^2 \partial_\xi \left\langle \frac{\Tilde{b}_i \Tilde{b}_i}{2} \right\rangle {+} \frac{1}{2} \partial_\xi \left\langle \Tilde{u}_3 \frac{\Tilde{b}_i \Tilde{b}_i}{2} \right\rangle {+} \frac{1}{2} \left \langle \Tilde{u}_i \partial_{\xi} \left( \frac{\Tilde{b}_{1} \Tilde{b}_i}{C_1} {+} \frac{\Tilde{b}_{2} \Tilde{b}_i}{C_2} {+} \Tilde{b}_{3} \Tilde{u}_i \right) \right \rangle {+} \frac{1}{2 \sqrt{\rho_m} \mathcal{A} g t} B_{0} \langle \Tilde{u}_i \partial_{\xi} \Tilde{b}_i \rangle {+} \frac{1}{2} \partial_{\xi} \langle \Tilde{b}_3 \Tilde{b}_i \Tilde{u}_i \rangle = 0.
    \end{split}
\end{equation}
We see that the magnetic field still deviates the self-similarity until the nonlinear terms dominate over the imposed magnetic field strength.

Similarly, the derivation of mixing layer height discussed in $\S$\ref{h-derivation} is independent of the orientation of the magnetic field and is hence expected to be valid for the vertical magnetic field as well. The value of the coefficients and hence $\alpha_{mhd}$ are required to be calculated by running numerical simulations.

\subsection{Relevance to HD RTI and comparison with other HD RTI studies}
The formula applies to both HD and magnetic field cases. It is not only a good tool to understand why $\alpha_{mhd}$ varies with magnetic field strengths, but also to understand why $\alpha_{hd}$ is different between studies. In table \ref{coefficients_comp}, we provide such a comparison. We notice that the $C_{diss}$ is lower for the current case relative to other studies. This could be due to differences in the numerical framework, grid resolution, and variance in parameters and initial perturbation between the present study and the other studies.  Several previous studies of HD RTI reported the energy dissipation, component-wise TKE in their studies. Based on the data provided, we calculate the $C_{diss}, C_{aniso}$ for their studies. The estimation of these coefficients from each study is different. For example, the values from \cite{Youngs1991} were calculated from Table II. The values from \cite{Cabot2006} were calculated by manually estimating the values between $t/\tau \in [12, 30]$ in figure 5. The dissipation and TKE values vary over time, and hence we provide a range for the coefficient. In the case of \cite{Stone2007a}, we calculate values based on Table II (specified at a particular mixing layer height), where we use the change in internal energy as a proxy for the total energy dissipated, as they do not include explicit dissipation in their numerical model. All these systematic errors lead to differences in the values of the coefficients. The other reason for the difference could also be due to the divergence-free velocity approximation, since the small-scale RTI mixing is highly sensitive to such assumptions. Note that $C_{ep}$ is zero for the HD RTI, since the TME is zero. 

We present the comparison of $\alpha_{hd}$ between the previous studies and the current study in Table \ref{coefficients_comp}. Note that the $\alpha_{hd}$ in table \ref{coefficients_comp} is reported based on bubble height $\left( (\alpha_{hd})_b = \frac{\alpha_{hd}}{1+|h_s/h_b|} \right)$ and not the full width of the mixing layer ($\alpha_{hd}$, as reported in figure \ref{alpha_comp}). $|\star|$ is the absolute value of $\star$. This is for the sake of comparison with other studies where $\alpha_{hd}$ is estimated based on bubble height.

\subsection{Effect of magnetic field strength and comparison with other works}

\begin{table}[H]
\begin{center}
\caption{Comparison of $C_{diss}, C_{ep}, C_{aniso}$ from different studies, including the present study. Note that here $(\alpha_{mhd})_b$ or $(\alpha_{hd})_b$ is reported based on bubble height ($h_b$) for comparison with other studies.}
\label{coefficients_comp}
\begin{tabular}{||c | c | c | c | c | c | c||} 
 \hline
 Paper & $B_0 (\% B_c)$ & A & $C_{diss}$ & $C_{ep}$ & $C_{aniso}$ & $(\alpha_{mhd})_b$ or $(\alpha_{hd})_b$ \\ [0.5ex] 
 \hline\hline
 \citet{Youngs1991} & 0 & 0.5 & 1.17 & 0 & 0.77 & 0.035 \\ 
 \hline
 \cite{Cabot2006} & 0 & 0.5 & $\approx1.31$ to $\approx0.97$ & 0 & 0.57 & $\approx$ 0.03 \\
 \hline
 \cite{Stone2007a} & 0 & 0.5 & 1.6 & 0 & - & 0.021 \\
 \hline
 This study & 0 & 0.5 & 0.946 & 0 & 0.58 & $\approx$ 0.0187 \\
 \hline
 This study & 1 & 0.5 & 0.93 & 0.027 & 0.55 & $\approx$ 0.02 \\
 \hline
 This study & 2 & 0.5 & 0.911 & 0.071 & 0.50 & $\approx$ 0.0203 \\
 \hline
 This study & 3 & 0.5 & 0.874 & 0.12  & 0.44 & $\approx$ 0.0234 \\
 \hline
 This study & 5 & 0.5 & 0.795 & 0.23  & 0.36 & $\approx$ 0.0262 \\
 \hline
 \cite{Stone2007a} & 10 & 0.5 & 1.1 & 0.53 & - & 0.031 \\
 \hline
 This study & 15 & 0.5 & 0.586 & 0.55 & 0.29 & $\approx$ 0.04423 \\
 \hline
 This study & 25 & 0.5 & 0.511 & 0.58 & 0.35 & $\approx$ 0.0447 \\
 \hline
\end{tabular}
\end{center}
\end{table}
While the $h$ grows quadratically in both HD and magnetic field cases, the magnetic field strength influences the growth of the layer by altering the above physical quantities (and thus $\alpha_{mhd}$). By simulating MRTI at different magnetic field strengths, we found that $C_{ep}$ and $C_{gr}$ increase with increasing field strength, $C_{diss}$ and $C_{aniso}$ decrease with increasing field strength. The variation of these quantities is particularly steep in the weak magnetic field case up to $5\%B_c$ (see table \ref{coefficients_comp}). This could be due to the significant changes in the turbulence. We compare the data from the current study with the \citet{Stone2007a}, which reports the energy dissipation, TKE, and TME. Similar to the HD case, the dissipation is much higher in \citet{Stone2007a}. The energy partition in the case of \cite{Stone2007a} is approximately as expected from the current study.
Overall, the value of $(\alpha_{mhd})_b$ was found to increase with increasing magnetic field strength, starting from $\approx 0.019$ for the HD case to $\approx 0.045$ for $B_0 = 25\%B_c$. The $(\alpha_{mhd})_b$ for the HD case is approximately the same for both \citet{Stone2007a} and this study (see table \ref{coefficients_comp}). In agreement with the trend reported by \cite{Stone2007a}, the value of $\alpha_{mhd}$ in this study increases with magnetic field strength. The steep increase in $C_{gr}$ (see figure \ref{C4}(right)) could be responsible for this. As the magnetic field strength increases, the turbulence in the mixing layer decreases, resulting in a significant proportion of vertical TKE invested towards the growth of the mixing layer, leading to an increase in $C_{gr}$ with magnetic field strength.

The results from the present study are consistent with \cite{Stone2007a} qualitatively --- decreasing $C_{diss}$ with increasing field strength, increasing $\alpha_{mhd}$ with increasing magnetic field strength. However, the values are very different due to the methodological differences. \cite{Stone2007a} used a code that is second-order accurate in space and time, at lower resolution, and with no explicit diffusion.

\subsection{Variable Density approximation}
We would like to reiterate that in the present work, we solve an approximation of the Variable Density problem, approximating velocity divergence to be zero. The sensitivity of RTI dynamics to velocity divergence criteria meant that it is possible that the approximation could lead to differences in small-scale mixing and consequently the nonlinear growth constant $\alpha_{mhd}$. However, we stress that, while the current assumption of $\nabla \cdot u = 0$ could change $\alpha_{mhd}$ and $C_{diss}$ quantitatively, it does not disprove our results of self-similarity, quadratic growth, and the validity of the derived analytical formula.

There is, however, an important discussion to have about the applicability of equation \ref{Cookdivu} to MHD. While the Fickian diffusion is valid for the HD case, the diffusion behaviour is expected to vary with the introduction of the magnetic field. The presence of a magnetic field introduces directionality in the system, and consequently into the key transport coefficients. A particle's random motion along a magnetic field line is driven by thermal motion, and hence we expect the particles to undergo Brownian motion in this direction, similar to the unmagnetized case. Thus, the diffusion along the magnetic field may effectively be similar to the HD case. Across the magnetic field, the particles undergo helical motion (gyromotion) that ties the particles to the magnetic field, making the system effectively immiscible across the magnetic field line unless this is broken. However, the coupling of the particles and the magnetic field is not perfect due to particle collisions. The introduction of particle collisions leads to non-zero resistivity that decouples the fluid and magnetic field, which is intrinsically related to electric currents in the system.

Thinking more about the particle motion, it is well known that particles have drift motion. Particle drift motions introduce cross-magnetic field drift, which may transport particles between the two fluid regions. The drift motion occurs due to complex field interactions like interaction between electric and magnetic field ($E \times B$ drift), gravitational and magnetic field ($g \times B$ drift), centrifugal forces, and magnetic field. Further, when the plasma flow becomes turbulent, it generates its own fluctuating electric and magnetic fields, which lead to particle drift across the magnetic field lines \citep{Bohm1949}. Thus, the mass diffusion across the magnetic field is predominantly due to particle collisions and drift motions, and not thermal Brownian motions of particles, making the Fickian diffusion potentially inappropriate across the magnetic field line. Thus, the mass diffusion in the presence of a magnetic field is very different \citep{Behram1962} from that of the miscible unmagnetized fluids studied in HD RTI so far. Following these arguments, the Fickian diffusion (and consequently equation \ref{Cookdivu}) is unlikely to be appropriate for the magnetic field case. 

Further research in this direction is essential to determine the appropriate diffusion formulation for the MHD case.It must be ensured that such a derived formulation would be match with the formulation of HD RTI in the limit of vanishing magnetic field, either in the cases of HD RTI simulations or null points in MHD simulations. The current formulation of $\nabla \cdot \textbf{u} = 0$ may not be necessarily satisfactory in MHD due to the fact that it does not lead to the variable density approximation for vanishing magnetic fields.

\subsection{Applicability to ICF:}

Note that in more complex systems where MRTI occurs, like ICF, the induction equation has additional terms due to the temperature and pressure gradients \cite{Walsh2017, Zhang_Jiang_Tao_Li_Yan_Sun_Zheng_2024}. In such systems, the magnetic field is evolved not only by the velocity, magnetic field coupling, but also self-generated from the baroclinic term (pressure gradient term). The applicability of this theory is uncertain. However, some predictions can be made in this context based on our current study. Since the pressure term was found to follow the self-similar behaviour, we expect the pressure gradient term in the induction equation could also have a self-similar fashion. The same is expected to be the case with the temperature term.

Besides, pressure and temperature gradient terms, there are collisional terms that diffuse magnetic fields. To understand their significance, we performed an analytical exercise assuming a simple Laplacian diffusion form ($D \nabla^2 \rho,$ $\nu \nabla^2 \mathbf{u},$ $\eta \nabla^2 \mathbf{b}$). The exercise showed that the diffusion terms do not comply with the assumed quadratic self-similar scaling. But, they decay much rapidly ($1/t^3$) compared to the imposed magnetic field term ($1/t$). Hence, relatively, the diffusion terms are expected to play an insignificant role in the self-similarity and growth of the mixing layer. Note that the simulations have explicit diffusion and solve non-ideal MHD governing equations.

\section{Conclusion}
The current study thus addresses several key gaps in the existing hydrodynamic and magnetohydrodynamic RTI. The key questions and answers are provided below.
\begin{enumerate}
    \item Is the HD self-similar scaling relevant to the MRTI? \newline We found that the HD scaling of mixing layer height ($h \propto t^2$), TKE $(\propto t^4)$ is relevant only when the system is well evolved that the influence of non-linear dynamics dominates that of the imposed magnetic field. The diminishing influence of the imposed magnetic field was analytically found to decay with time as $1/t$. In the eventual self-similar regime, we found the mixing layer grows as $h \approx \alpha_{mhd} \mathcal{A} g t^2$ for $t \gg 1$.
    \item How does the magnetic field influence the evolution of the mixing layer? \newline The first step to answer this question is to understand the factors that influence the non-linear growth constant ($\alpha_{mhd}$), which is lacking so far. Based on energy conservation, we found that $\alpha_{mhd}$ is a function of --- $i)$ energy partition between the kinetic energy, magnetic energy, and energy dissipation, and $ii)$ the growth rate of the mixing layer for unit vertical kinetic energy. These processes were used to understand the variation of $\alpha_{mhd}$ across different magnetic field strengths and compare the $\alpha_{mhd}$ between different studies. We found that the suppression of small-scale structures by the magnetic field led to increased anisotropy, which further improved the growth rate of $h$ per unit vertical kinetic energy. Thus, the increase in magnetic field contributes to an increase in $\alpha_{mhd}$. The $\alpha_{mhd}$ values are in good agreement with the MRTI study by \cite{Stone2007a}. Further, we found that the energy partition between dissipation, kinetic energy, and magnetic energy scales quadratically with imposed magnetic field in the weak magnetic field strength $(B_0 < 5\%B_c)$. However, this scaling becomes inappropriate in the regime where the mixing layer is dominated by the magnetic field and consequently has only large-scale plumes.
\end{enumerate}

\section*{Acknowledgements}
For the purpose of open access, the author has applied a ‘Creative Commons Attribution (CC BY) licence to any Author Accepted Manuscript version arising from this submission. AH would like to acknowledge support by the Research Institute for Mathematical Sciences, an International Joint Usage/Research Center located in Kyoto University. AH would like to thank Dr Takeshi Matsumoto for the insightful discussion on Rayleigh-Taylor mixing. The first author would particularly like to thank the anonymous Reviewer 2 for the insightful comments and discussion, the efforts they took to ensure the quality of the work, which helped improve the manuscript significantly. 

\vspace{-10pt}
\section*{Funding}
MTK is supported by the Engineering and Physical Sciences Research Council (EPSRC) Grant No. EP/W523859/1. AH is supported by STFC Research Grant No. ST/R000891/1 and ST/V000659/1. 
This work used the DiRAC Memory Intensive service (Cosma7) at Durham University, managed by the Institute for Computational Cosmology on behalf of the STFC DiRAC HPC Facility (www.dirac.ac.uk). The DiRAC service at Durham was funded by BEIS, UKRI and STFC capital funding, Durham University and STFC operations grants. DiRAC is part of the UKRI Digital Research Infrastructure capital grant ST/K00042X/1, STFC capital grant ST/K00087X/1, DiRAC Operations grant ST/K003267/1 and Durham University. DiRAC is part of the National E-Infrastructure.

\vspace{-10pt}
\section*{Declaration of interests}
The authors report no conflict of interest.
\vspace{-10pt}
\section*{Data availability statement}
The data that support the findings of this study are available from the corresponding author upon reasonable request.

\printcredits

\bibliographystyle{cas-model2-names}

\bibliography{cas-refs}

\end{document}